\documentclass[preprint,11pt]{JHEP3}

\JHEPspecialurl{http://jhep.sissa.it/JOURNAL/JHEP3.tar.gz}

\usepackage{epsfig,multicol,amsmath}

\usepackage{epstopdf}
\usepackage{bm}  
\usepackage{amsmath}     
\usepackage{epsfig,epsf}
\usepackage{graphicx}
\usepackage{rotating}
\usepackage{cite}

\def\beq{\begin{equation}}
\def\eeq{\end{equation}}
\def\bea{\begin{eqnarray}}
\def\eea{\end{eqnarray}}

\def\eq#1{{Eq.~(\ref{#1})}}
\def\fig#1{{Fig.~\ref{#1}}}
\newcommand{\bas}{\bar{\alpha}_S}
\newcommand{\as}{\alpha_S}

\newcommand{\Lb}{\left(}
\newcommand{\Rb}{\right)}
\parskip=\itemsep               

\newcommand{\nn}{\nonumber}

\newcommand{\h}{\frac{1}{2}}

\newcommand{\ga}{\gamma}

\newcommand{\pom}{I\!\!P}

\def\pom{{I\!\!P}}

\title{ A model for strong interactions at high energy based on 
the CGC/saturation approach}
\author{\Large  E. Gotsman$^{a}$\thanks{Email:
gotsman@post.tau.ac.il.}\,, E. Levin$^{a,b}$\thanks{Email:
leving@post.tau.ac.il}\,\,and\,\,U. Maor$^{a}$\thanks{Email: maor@post.tau.ac.il.}\, 
\\
a)\,  \,Department of Particle Physics, School of Physics and Astronomy,
Raymond and Beverly Sackler
 Faculty of Exact Science, Tel Aviv University, Tel Aviv, 69978, Israel\\ 
b)\,\,Departamento de F\'\i sica, Universidad T\'ecnica Federico Santa 
Mar\'\i a, Avda. Espa\~na 1680\\ and Centro 
Cientifico-Tecnol$\acute{o}$gico de Valpara\'\i so,Casilla 110-V, 
Valparaiso, Chile

\\}

\abstract{  We present our first attempt to develop a 
  model  for  soft interactions at high energy, 
based on the BFKL Pomeron and the CGC/saturation 
  approach.
 We construct  an  eikonal-type model, whose opacity  is determined by the
 exchange of the dressed BFKL Pomeron.
 The
 Green's function of the Pomeron
 is calculated in the framework of the CGC/saturation approach. Using 
 five parameters we  achieve a   reasonable description of the 
experimental data 
at
 high energies ( $W\,\geq\,0.546\,TeV$)  with overall
 $\chi^2/d.o.f. \approx 2$.  The model results in
 different behaviour for the  single and double diffraction cross 
sections  
at high energies. The single diffraction cross section  reaches a 
saturated value
 (about 10 mb) at high energies, while the double diffraction cross 
section continues growing slowly.}

\dedicated{PACS: 13.85.-t, 13.85.Hd, 11.55.-m, 11.55.Bq}
\preprint{TAUP - 2985/14\\
{\tt }\\
\today}
\begin{document}

\section{Introduction}
The strong interaction at high energies is one of the most difficult
 and unrewarding problems of  high energy physics. The reason  for this,
  is  the embryonic stage  of our understanding of 
 non-perturbative QCD. Traditionally, we consider the strong
 interaction at high energy as a typical example of  processes
 that occur at long distances, where the unknown force confining  
quarks
 and gluons plays a crucial role, making all our theoretical efforts to
 treat these processes, fruitless.   The description of these
 processes  which we need for practical purposes, is the 
field
 of high energy phenomenology, based on  Pomeron calculus 
\cite{COL,SOFT,LEREG}.
The LHC data\cite{ALICE,ATLAS,CMS,TOTEM} shows  that in many cases   
models based
 on this phenomenology failed to agree  with  the results of the 
classical 
set of soft interaction data: total, elastic and diffraction cross section as
 well as elastic slope and the inclusive  production of the secondary hadrons
 \cite{DL,GLM1,GLM2,KAP,KMR,OST}.
  However, there is a glimpse of hope due to the following  two facts:
 models that fit  the LHC data have been proposed 
\cite{NEWGLM,NEWKMR,OSTREC};
 and after two decades of  experience in  high energy phenomenology 
 we have learned, that the more 
  theoretically based the  phenomenological input is,
the more appropriate and  apprehensible, the resulting  description of the 
data we obtain.

In Ref.\cite{GLMREV} we reviewed our model which   describes successfully  
all high
 energy data, including those at the LHC, and which incorporates 
 theoretical ingredients from  N=4 SYM\cite{BST,POST,BFKL4}
 and from perturbative QCD\cite{BFKL,LI,BART}.  In the present paper we
  improve this approach, by including more pertinent theoretical
 input. First, we  introduce a more constructive meaning to     
 our old idea \cite{GLMONEPOMERON,GLM2}, that there is  only one
 Pomeron that describes both soft and hard interactions.  In perturbative
 QCD the BFKL Pomeron at high energy  takes the 
following
 form \cite{BFKL}\footnote{
\eq{I1} gives the Pomeron contribution at high energies for
 $\ln^2 w w^*  \leq 4 D Y$. For a  wider kinematic region the contribution
 of the BFKL Pomeron  can be found in Ref.\cite{LI}.}

\bea \label{I1}
&&G_\pom\Lb Y, r, R; b\Rb\,\,=\,\,\Lb w\,w^* \Rb^\h \sqrt{\frac{\pi}{4\,D\,Y}}\,e^{ \Delta_{\mbox{\tiny BFKL}}\,Y \,-\,\frac{\ln^2 w\,w^*}{4 \,D\,Y}}\\
&&\mbox{with}~~\Delta_{\mbox{\tiny BFKL}}\,=\,2 \ln 2\,\bas~\mbox{and} ~D\,
=\,14 \zeta(3) \bas \,=\,16.828 \,\bas\nn .
\eea
  $G_\pom\Lb Y, r, R; b\Rb$ denotes the BFKL Pomeron Green
 function,  $\bas$  the QCD coupling,  $r$ and $R$ are  the
 sizes of two interacting dipoles. $Y \,=\,\ln s$, where
 $s = W^2$.  $W$ denotes the energy of the interaction, and $b$  the 
impact 
parameter for the scattering amplitude of two dipoles.  

\beq \label{W}
w\,w^*\,\,=\,\,\frac{r^2\,R^2}{\Lb\vec{b} - \h\Lb\,\vec{r}\,
- \,\vec{R}\Rb\Rb^2
\,\Lb\vec{b} \,+\, \h \Lb\,\vec{r} \,- \,\vec{R}\Rb\Rb^2}.
\eeq
From \eq{I1} it is obvious that 
the BFKL Pomeron is not  a pole in   angular 
momentum, but a branch cut, since its Y-dependence  
has an additional $\ln s$ term; it does not reproduce the exponential
 decrease at large $b$, which follows from the general properties of
 analyticity and unitarity \cite{FROI}; as  the exchange of the BFKL 
Pomeron
depends on the sizes of the dipoles, consequently, 
the BFKL Pomeron does not factorize.

  The Pomeron that appears in N=4 SYM \cite{BST},
 corresponds to the BFKL Pomeron in QCD with the following glossary:
\par
\begin{center}
{\bf  Glossary~~~ ${ \equiv}$\,\,~~~ AdS-CFT correspondence\cite{AdS-CFT}:}
~
\begin{tabular}{  c c c}
N=4 SYM &    & { QCD}\\
&~~~~~~~~~~~ & \\
{Reggeized graviton} \,&${ \iff}$&\,{ BFKL Pomeron}\\
{$z $} \,&${ \iff}$ & $\,r $\,(dipole size)\\
$ 1 - 2/\sqrt{\lambda}$ & $ \iff$&\,$ \Delta_{BFKL}$ 
(intercept of the BFKL Pomeron)\\
$2/\sqrt{\lambda} $\,&$ \iff$&\, $D$ \\
\end{tabular}
\end{center}
where $\lambda = 4 \pi N_c \alpha^{YM}_S$ and $\alpha^{YM}_S $ denotes  
the QCD-like 
coupling.

Hereby, we generalize our approach, dealing with the BFKL Pomeron,
 instead of the simple Regge pole that was used in our previous model.

The second innovation is related to the Pomeron interaction. 
 The LHC data supports the assumption that the dense system of
 partons (gluons) are produced in the proton-proton interaction at
 high energy. Such a system of partons appears naturally  in the 
CGC/saturation
 approach \cite{GLR,MUQI,MV,MUCD,BK,JIMWLK,REV},  and provides a 
successful 
description 
of the general properties of the average  event at the LHC\cite{LERE}, 
and of the
 long range rapidity angular correlations \cite{COR}. In this paper, we
 use the CGC/saturation approach to describe the Pomeron interactions,
  replacing   the Pomeron calculus. This strategy allows us, not only to  
treat
 the Pomeron interactions, but also to include the  saturation phenomenon, 
which was beyond the scope of our previous model.


\section{ Model:  theoretical ingredients}



\subsection{Parton cascade of one dipole  in the saturation region}
\begin{figure}[ht]
    \centering
  \leavevmode
      \includegraphics[width=13cm]{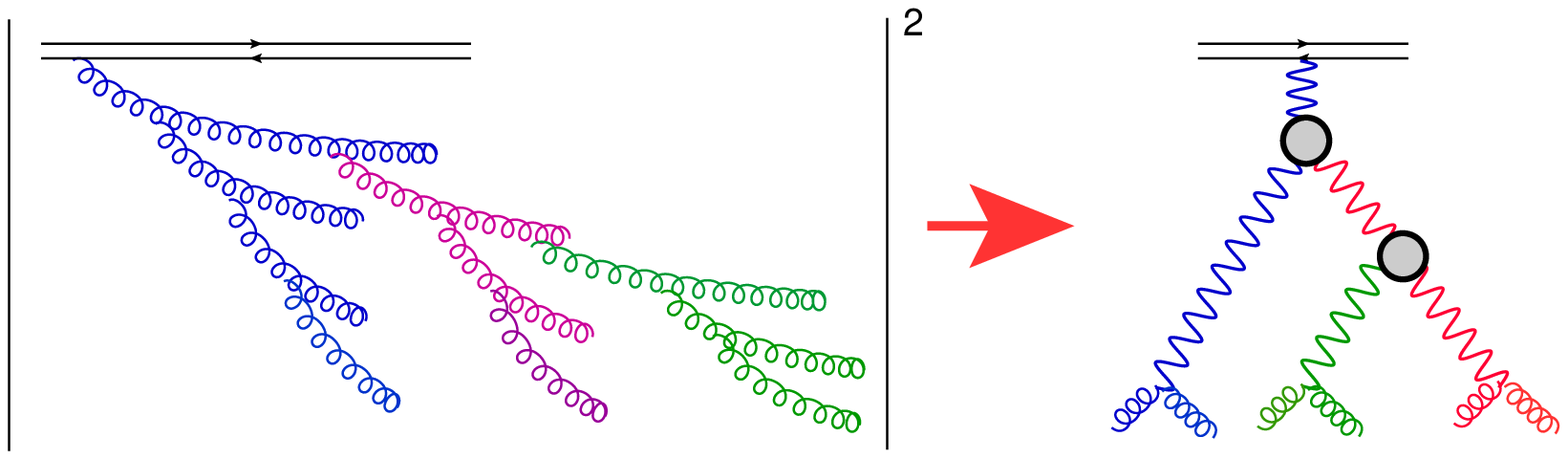}  
      \caption{The structure of the parton cascade for a fast dipole and 
its relation to the Pomeron interaction. Helical lines denote gluons.
 The wavy lines describe  BFKL Pomerons. The blobs stand for
 triple Pomeron vertices.}
\label{cascade}
   \end{figure}

The parton cascade which  originates from the decay of  one gluon
 to two gluons in QCD, can be described equally well in two ways.
 The first, using
 the QCD expression for this decay,  describes the change of  probability 
to have $n$ gluons
 at rapidity $Y$, due to the decay of one gluon to two. 
   The equation in this
 approach is a linear functional equation for the generating
 functional (see Refs. \cite{MUCD,JIMWLK,LELU}). The alternate way
 is to sum the Pomeron  fan diagrams (see \fig{cascade}) in the
 framework of the BFKL Pomeron calculus\cite{BRN}.

 In this paper we  use the solution of the functional equation
 which  was proposed and discussed  in Ref.\cite{LEPP}.  
   For  completeness of presentation,
we  repeat  the main ideas of the solution, and explain the
 physical meaning of the phenomenological parameters that we have
  introduced in our model.
 
 The first simplification arises when we consider the interaction of the parton 
cascade with a large target (say with a heavy nucleus). In
 this case the functional equation reduces to the non-linear Balitsky-Kovchegov
 equation. The solution of this equation has three distinct kinematic regions. 
 \begin{enumerate}
 \item $r^2 Q^2_s\Lb Y, b\Rb \,\ll\,1$, where $Q_s$ denotes the saturation 
scale\cite{GLR,MUQI,MV}. The non-linear corrections are small, and the
 solution is the BFKL Pomeron;
  \item $r^2 Q^2_s\Lb Y, b\Rb \,\sim\,1$ (vicinity of the saturation scale).
 The scattering amplitude has the following form\cite{MUT,IIM}
  \beq \label{TI1}
   A\,\,\equiv\,\,G_\pom\Lb \phi_o, z\Rb\,\,=\,\,\phi_0 \Lb r^2 \,
 Q^2_s\Lb Y, b\Rb\Rb^{1 - \gamma_{cr}},
   \eeq 
 where $\phi_0$ is a constant, and  where the critical anamolous 
dimension $\gamma_{cr}$, can be found from
 \beq \label{GACR}
\frac{\chi\Lb \gamma_{cr}\Rb}{1 - \gamma_{cr}}\,\,=\,\, -
 \frac{d \chi\Lb \gamma_{cr}\Rb}{d \gamma_{cr}}~~~\,\,\,\mbox{and}\,\,\,
~~~\chi\Lb \gamma\Rb\,=\,\,2\,\psi\Lb 1 \Rb\,-\,\psi\Lb \gamma\Rb\,-\,\psi\Lb
 1 - \gamma\Rb.
\eeq
 \item $r^2 Q^2_s\Lb Y, b\Rb \,\gg\,1$ (deeply inside the saturation domain).
 The amplitude  approaches unity\cite{LT}: viz.
   \beq \label{TI2}
   A\,\,=\,\, 1\,\,-\,\,\mbox{Const}\,\exp\Big(  - \frac{z^2}{2\,\kappa}\Big)
  , ~~\mbox{where}~~z\,\,=\,\,\ln\Lb r^2\,Q^2_s\Lb Y, b\Rb\Rb\,\,\,
 =\,\,\,\,\bas\,\kappa \,Y\,\,+\,\,\xi\,~\mbox{,}~\kappa\,=
\,\,\,\frac{\chi\Lb \gamma_{cr}\Rb}{ 1 - \gamma_{cr}}.   
      \eeq
 and  $ \xi\, = \,\ln \Lb r^2\,Q^2_s\Lb Y\,=\,Y_{0}, b\Rb\Rb $ .    
      \end{enumerate}
    
     In spite of our understanding of all qualitative features of the
 solution, we do not have an  equation for an analytical solution 
\cite{REV},
 which we need to reconstruct the parton cascade. The parton cascade can be
 described as the amplitude for the production of dipoles of size 
$r_i$ at
  impact parameters $b_i$.  This amplitude can be 
written as
    \bea \label{TI3}
&&N\Lb Y- Y', r, \{ r_i,b_i\}\Rb\,\,=\\
&&\,\,\sum^{\infty}_{n=1} \,\Lb - \,1\Rb^{n+1} \widetilde{C}_n\Lb
 \phi_0, r\Rb \prod^n_{i=1} G_\pom\Lb Y - Y';  r, r_i , b_i\Rb\,\,
=\,\,\sum^{\infty}_{n=1} \,\Lb - \,1\Rb^{n+1} \widetilde{C}_n\Lb
 \phi_0, r\Rb \prod^n_{i=1} G_\pom\Lb z - z_i\Rb\nn.
\eea
 
 The solution to the non-linear equation is of the following general form
 \beq \label{TI4}
N\Lb G_\pom\Lb \phi_0,z\Rb\Rb \,\,=\,\,\sum^{\infty}_{n=1} \,\Lb - 
\,1\Rb^{n+1} C_n\Lb \phi_0\Rb G_\pom^n\Lb \phi_0,z\Rb.
\eeq   
Comparing \eq{TI3} with \eq{TI4} we see 
\beq \label{TI5}
\widetilde{C}_n\Lb \phi_0, r\Rb\,\,\,=\,\,\,C_n\Lb \phi_0\Rb.
\eeq 
Unfortunately, we cannot find the coefficient  $C_n$, for the general 
non-linear
 equation.  For the case of the simplified BFKL kernel (see Refs. 
\cite{LT,LEPP})
 the solution  can be found, and  we can suggest a simple formula that 
provides a very accurate
  solution of \eq{TI4}  (see Ref.\cite{LEPP}).
\beq \label{TI6}
N\Lb G_\pom\Lb \phi_0,z\Rb\Rb \,\,=\,\,a\,\Lb 1
 - \exp\Lb -  G_\pom\Lb \phi_0,z\Rb\Rb\Rb\,\,+\,\,\Lb 1 - a\Rb
\frac{ G_\pom\Lb \phi_0,z\Rb}{1\,+\, G_\pom\Lb \phi_0,z\Rb},
\eeq
with $a$ = 0.65. 

This formula allows us to find $\widetilde{C}_n\Lb \phi_0, r\Rb$,
 and to reconstruct the amplitude of \eq{TI3}.
\subsection{Summing Pomeron loops (MPSI approximation)}
\begin{figure}[ht]
    \centering
  \leavevmode
      \includegraphics[width=13cm]{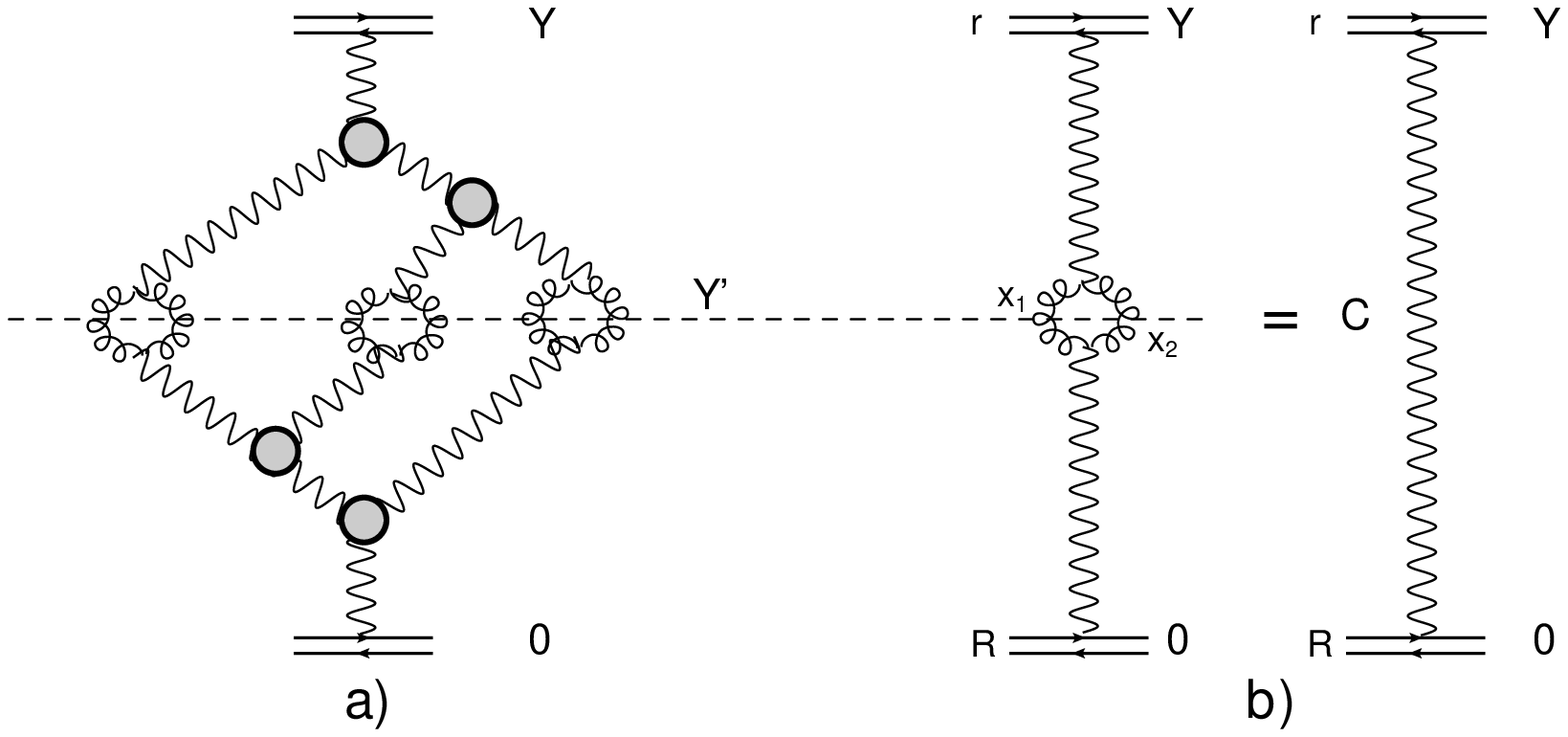}  
      \caption{ MPSI approximation: the simplest diagram(\protect\fig{mpsi}-a)
 and  one Pomeron contribution (\protect\fig{mpsi}-b).
 $C\,\,=\,\,\bas^2/4 \pi$. $ \vec{x}_1 - \vec{x}_2  =
 \vec{r}^{\,'}$. $\h (\vec{x}_1 + \vec{x}_2 ) = \vec{b}^{\,'}$.
  The wavy lines describe  BFKL Pomerons.
 The blobs stand for triple Pomeron vertices.}
\label{mpsi}
   \end{figure}

It was shown in Ref.\cite{LMP}, that in the BFKL Pomeron calculus
  for the parton cascade (see \fig{cascade}), the integration over
 rapidities of the triple Pomeron vertices, suggests that the value of the
 typical rapidity is of the order of $Y - Y_i \sim 1/\Delta_{\mbox{\tiny 
BFKL}}$.
  Consequently, only large Pomeron
 loops with rapidity    of  order  $Y$, contribute at high energies
 \cite{MPSI}. To sum such loops we use the MPSI approximation developed
 in Ref.\cite{MPSI}. The essence of this approximation is to use the 
$t$-channel
 unitarity constraint, which is satisfied by the one BFKL Pomeron 
exchange. 
Indeed,  at any value of $Y'$, 
 the BFKL Pomeron 
has the following property from  $t$-channel unitarity
 \cite{GLR,MUSH} (see \fig{mpsi}-b)
\beq \label{POMTUN}
\frac{\as^2}{4 \pi} \,\,G_\pom\Lb Y - 0, r, R; b  \Rb\,=\,\int d^2 r'
  d^2 b' \,G_\pom\Lb Y - Y', r, r', \vec{b} - \vec{b}^{\,'} \Rb \, \,
\,G_\pom\Lb Y' r', R, \vec{b} - \vec{b}^{\,'} \Rb\,\,.
\eeq
The MPSI approximation is illustrated in \fig{mpsi}-a,  where the first
 non-trivial loop diagram is presented.  This approximation
 enables us to evaluate the Pomeron loops, using the fan diagram structure 
of
 the parton cascade.  The general MPSI equation for the sum of
 enhanced Pomeron  diagrams,  has the form    which leads to a 
new Pomeron Green function (dressed Pomeron).
\bea \label{MPSI1}
G^{\mbox{dressed}}_\pom\Lb Y, r, R; b \Rb\,\,&=&\,\,\int \prod_{i =1} d^2 r_i\, d^2 b_i\, N\Lb Y- Y', r, \{ r_i,b - b_i\}\Rb\,
N\Lb  Y', R, \{ r_i,b_i\}\Rb\nn\\
&=&\sum^{\infty}_{n=1} \,\frac{\Lb - \,1\Rb^{n}}{n!} \,\Lb \frac{\bas}{4\pi}\Rb^n\Lb \widetilde{C}_n\Lb \phi_0, r\Rb \widetilde{C}_n\Lb \phi_0, R\Rb\Rb^n\,  G^n_\pom\Lb Y - 0;  r, R, b \Rb\nn\\
&=& \sum^{\infty}_{n=1} \,\frac{\Lb - \,1\Rb^{n}}{n!}
 \,\Big( \frac{\bas}{4\pi}\, C^2_n\Lb \phi_0\Rb\Big)^n\,
  G^n_\pom\Lb z \Rb .\eea

In the last equation we used \eq{TI5} with
\beq \label{Z}
z\,\,=\,\,\ln\Big( r^2\,Q^2_s\Lb Y, R; b\Rb\Big).
\eeq
Since, for the proton-proton scattering, $r = R $, $\, z \,>\,0$  
(see 
\eq{Z}), 
  we are
 dealing with  parton cascades in the saturation domain.
  We recall that the saturation domain corresponds to
 $ z>0, \,( r^2 Q^2_s \,\geq\,1$), while $z < 0 \,( r^2Q^2_s
 \,\leq\,1)$ characterizes the perturbative QCD region.

In Ref.\cite{AKLL} the MPSI approximation, as well as the equivalence of
 the CGC/saturation approach and the BFKL Pomeron calculus, was proven for 
a  wide range of rapidities:
\beq \label{MPSI2}
Y\,\,\leq\,\,\frac{2}{\Delta_{\mbox{\tiny BFKL}}}\,\ln\Big(\frac{1}
{\Delta^2_{\mbox{\tiny BFKL}}}\Big).
\eeq

For larger $Y$  the MPSI approximation does not give the exact answer,
  since we have not   introduced the vertex of the four Pomeron 
interaction,
 which violates the simple structure of the parton cascade shown in
 \fig{cascade}.
The errors that stem from neglecting the four Pomeron interaction, have 
been
 evaluated in Ref.\cite{GLMREV,AKLL}.

\section{Model: main formulae and parameters}
In this section we  describe our model.  Its  main ingredient
 is the sum of the Pomeron loops, that leads to a new dressed Pomeron 
Green function.

\subsection{Dressed Pomeron}
The resulting Green function of the Pomeron is given by
 \eq{MPSI1}. Using \eq{TI6} and \eq{MPSI1} we  obtain
 the following expression (see Refs.\cite{LEPP,GLMREV} for more details ):
\bea\label{DP1}
&&G^{\mbox{dressed}}_\pom\Lb Y- Y_0, r,  R, b \Rb\,\,=\,\, \\
&&a^2\Bigg\{ 1 \,-\,\exp\Lb - T\Lb Y- Y_0, r,  R, b \Rb \Rb \,\Bigg\}\,+\,2 a (1 -  a) \frac{ T\Lb Y- Y_0, r,  R, b \Rb }{
1\,\,+\,\,T\Lb Y- Y_0, r,  R, b \Rb }\nn\\
  && +\,\,( 1 -  a)^2 \,\Bigg\{1 - \exp\Lb \frac{1}{ T\Lb Y- Y_0, r,  R, b \Rb }\Rb\,\frac{1}{T\Lb Y- Y_0, r,  R, b \Rb }\,
  \Gamma\Lb 0,   \frac{1}{
T\Lb Y- Y_0, r,  R, b \Rb }\Rb   \Bigg\}\nn ,
 \eea 
 where $\Gamma\Lb  x \Rb$ is the incomplete Euler
 gamma function (see {\bf 8.35} of Ref.\cite{RY}). The
 function $T\Lb Y- Y_0, r,  R, b \Rb$ can be found
 from \eq{MPSI1} and has the form:
 \beq \label{DP2}
 T\Lb Y- Y_0, r,  R, b \Rb \,\,=\,\,\frac{\bas^2}{4 \pi}\,G_\pom\Lb
 z \to 0 \Rb \,\,=\,\,\phi_0 \Lb r^2 Q^2_s\Lb R, Y,b \Rb\Rb^{1 -
 \gamma_{cr}}\,=\,\phi_0 S\Lb b \Rb e^{\lambda (1 - \gamma_{cr}) Y},
 \eeq
 where we used two inputs: $r = R$ and $Q^2_s\,=\,\Lb 1/(m^2R^2)\Rb\,S\Lb 
b \Rb
 \, \exp\Lb \lambda \,Y\Rb$.
  The parameter  $\lambda\,=\,\bas
 \chi\Lb \gamma_{cr}\Rb/(1 -
 \gamma_{cr})$ in  leading order of perturbative QCD.  From 
phenomenology
 $\lambda $ turns out to have the value $\lambda = 0.2 \div 
0.3$\cite{GW,KLN,AAQ}.
 $S\Lb b\Rb$ is a pure phenomenological profile function which
 we choose to be of the
 form
  \beq \label{S}
  S\Lb b \Rb\,\,=\,\,\frac{m^2}{2 \pi} \,e^{- m\,b},~~~~~~~~~~~~\mbox{with 
normalization}~~~\int d^2 b \,S\Lb b\Rb\,=\,1 .
  \eeq
The parameter $m$ represents  the inverse size of the dipole $m \sim 1/r = 
1/R$. 
Unfortunately, we have no theoretical estimate for this mass.  It
 maybe large, reflecting the masses  of glueballs and the small size
 of the typical dipole in a hadron \cite{SN}. Note, that $S\Lb b\Rb$
 has a correct, exponential decrease at large $b$. This is an advantage
 of our approach, as it enables us to  introduce a non-pertutbative scale 
in a
 physical motivated way, for the observable
that  characterizes the principle property of the parton cascade. Therefore,
 in the framework of our approach we do not face the theoretical problem of
 large $b$ behaviour, which is the main unsolved problem in 
the CGC/saturation
 approach \cite{KOWI}.

     \begin{figure}[ht]
    \centering
  \leavevmode
      \includegraphics[width=13cm]{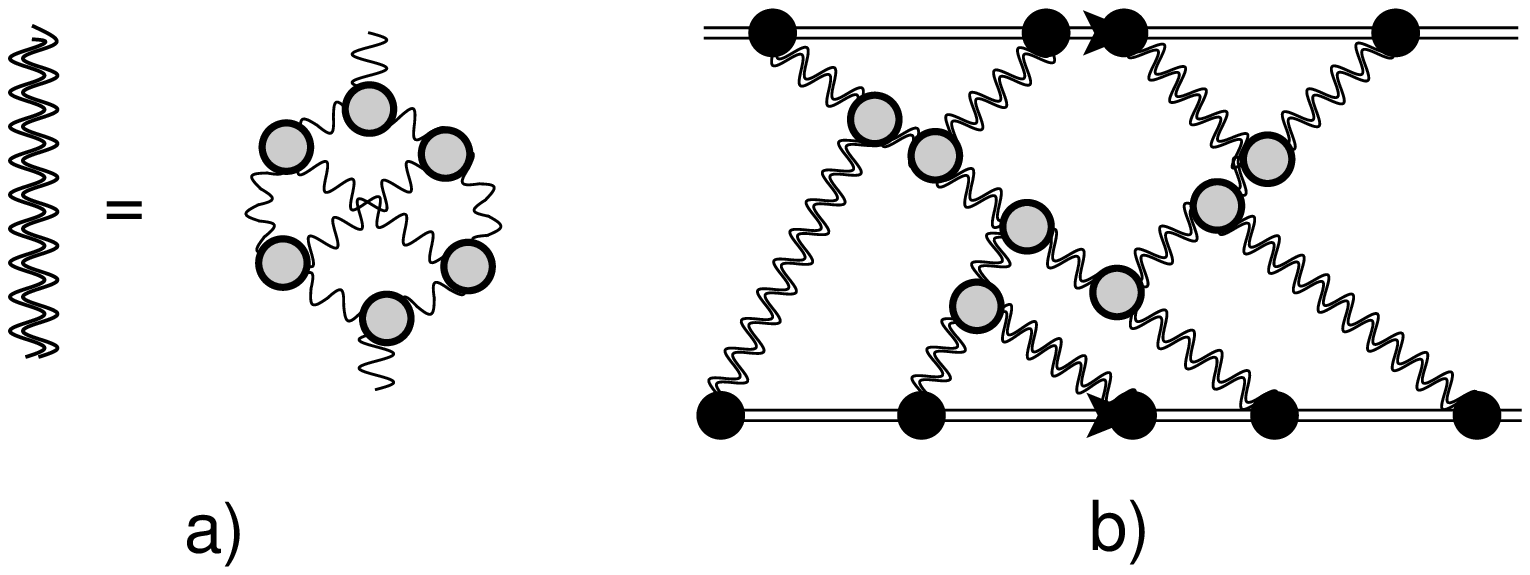}  
      \caption{  The dressed Pomeron in the MPSI approximation (\protect\fig{netdi}-a) and the sum of net diagrams (\protect\fig{netdi}-b).
 The wavy lines describe  BFKL Pomerons. The gray  blobs stand
 for triple Pomeron vertices while the black ones describes the vertex
 hadron-Pomeron interaction $g\Lb b \Rb$.}
\label{netdi}
   \end{figure}

 
\subsection{Interaction of dressed Pomerons}

The interaction of a dressed Pomeron with a hadron is a non-perturbative
 problem, which cannot be solved at the moment.  From the microscopic 
point 
of
 view this interaction depends on the size of a typical dipole in a 
hadron,
  on the probability of finding such a dipole, and on the interaction 
coupling.
 Since this interaction  originates at long distances, we cannot calculate 
it
  even in the CGC/saturation approach.  Introducing two phenomenological
 constants: $g$ and $m_1$,  we describe the vertex of the  hadron-Pomeron 
interaction as follows
\beq \label{ND1}
g\Lb b\Rb\,\,=\,\,g\,S_h\Lb b \Rb\,\,=\,\,\mbox{with}\,
~~S_h\Lb b \Rb\,\,=\,\,\frac{m_1^3\,b}{4 \pi}\,K_1\Lb m_1 b \Rb,
~~~\mbox{where}~~ S_h\Lb b \Rb\,\xrightarrow{\mbox{Fourier image}}\,~
 (\frac{m_1^2}{q^2\,+\,m_1^2})^{2}.
\eeq 

To account for the interaction of the dressed Pomerons  with hadrons, we
 use the strategy that has been suggested in Ref.\cite{GLMMA}, and  which
 is based on the fact that we anticipate   the value of $g$ in 
\eq{ND1} 
will
 be large.  In this case we can  evaluate the scattering amplitude
 in the following kinematic region of rapidities:
\beq \label{ND2}
g\,G_{3\pom} \,G^{dressed}_\pom\Lb Y, b=0\Rb\,\,\approx\,\,1; ~~~~~~\mbox{while}~~~~~
\,G^2_{3\pom} \,G^{dressed}_\pom\Lb Y, b=0\Rb\,\,\ll\,\,1.
\eeq
The difference with our previous model reviewed in Ref.\cite{GLMREV} 
 lies in the value of $G_{3 \pom}$ which was a phenomenological parameter,
 while now we are  able to estimate it from the CGC/saturation approach.

Finally, the opacity $\Omega$ has the form
\beq \label{ND3}
\Omega\Lb Y; b\Rb~~= ~~ \int d^2 b'\,
\,\,\,\frac{ g\Lb\vec {b}'\Rb\,g\Lb\vec{b} - \vec{b}'\Rb\,\bar{G}^{\mbox{dressed}}_\pom \Lb Y\Rb
}
{1\,+\,1.29\,\bar{G}^{\mbox{dressed}}_\pom \Lb Y\Rb\Big[
g\Lb\vec{b}'\Rb + g\Lb\vec{b} - \vec{b}'\Rb\Big]} ,
\eeq
where 
\beq \label{ND4}
\bar{G}^{\mbox{dressed}}_\pom \Lb Y\Rb\,\,=\,\,\int d^2\,b'
 \,G^{\mbox{dressed}}_\pom \Lb Y; b' \Rb.
\eeq

In \eq{ND3} we assumed that $m\,\gg\,m_1$.  The factor
 $1.29$ stem from  estimates of the triple Pomeron
 vertex in the CGC/saturation approach.

\subsection{Observables}

\subsubsection{Elastic amplitude}
The elastic amplitude is

\bea \label{O1}
A_{el}\Lb Y; b \Rb~~&=& ~~i \Big( 1\,\,-\,\,e^{ - \Omega\Lb Y,b\Rb}\Big).
\eea
\subsubsection{Single diffraction}
\begin{figure}[ht]
    \centering
  \leavevmode
      \includegraphics[width=13cm]{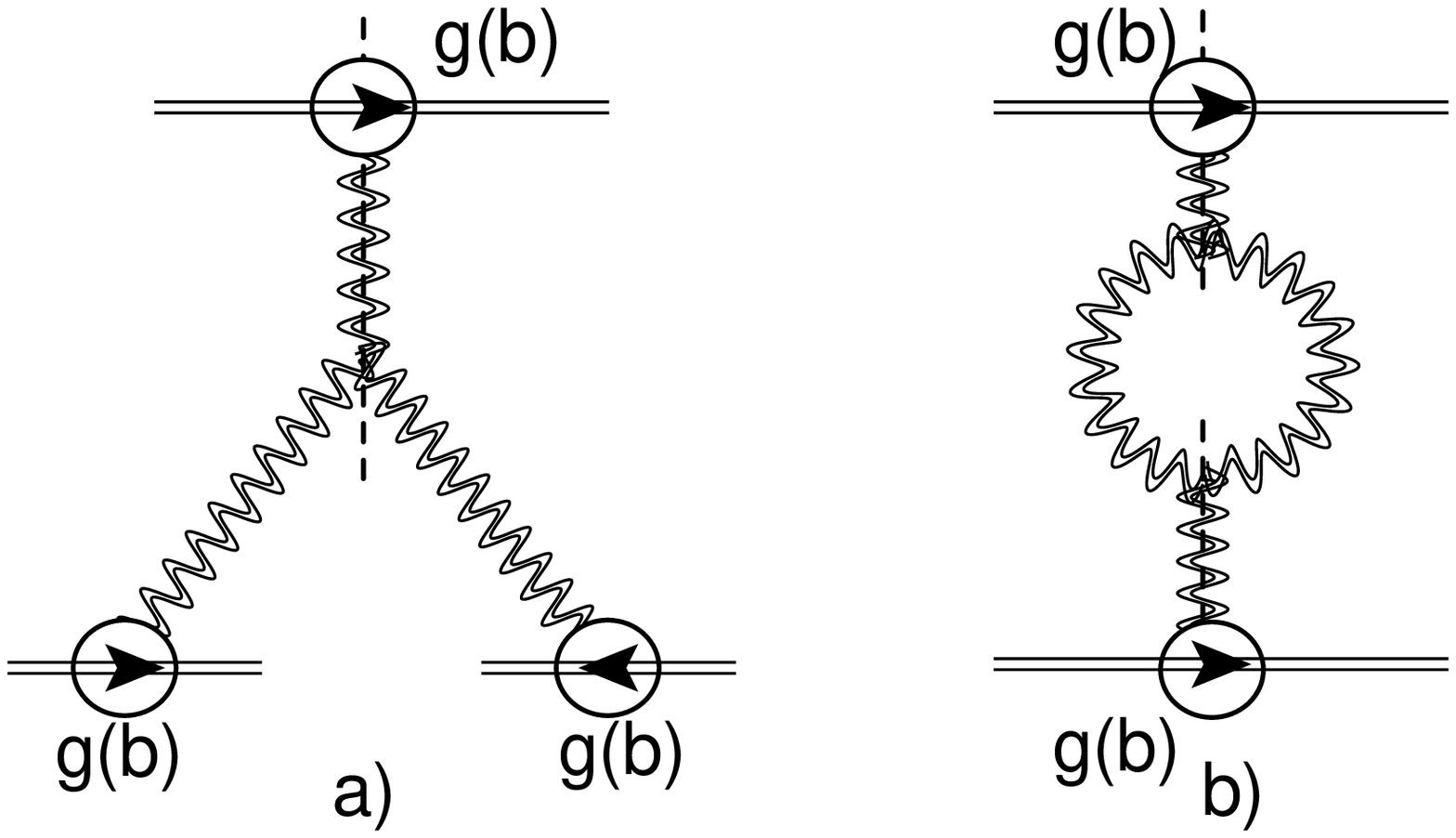}  
      \caption{  Single (\protect\fig{sddd}-a) and double 
(\protect\fig{sddd}-b) diffraction production.
The double wavy lines describe  dressed Pomerons.
 The double wavy line crossed by the dashed one  stands
 for the dressed Pomeron structure, in terms of produced particles. 
The  blobs stand for the  hadron-Pomeron interaction $g\Lb b 
\Rb$ vertices.}
\label{sddd}
   \end{figure}

The cross section for single diffraction can be written as
\beq \label{O2}
\sigma_{SD}\Lb Y \Rb\,\,=\,\,2\,\int d^2 b \frac{d \sigma_{SD}}{d b^2}\,\,=\,\,\int d^2\, b g^3\,S_{SD}
\Lb b \Rb \bar{N}_{SD}\Lb Y\Rb\,e^{ - 2\,\Omega\Lb Y,b\Rb},
\eeq
where $ \bar{N}_{SD}\Lb Y\Rb\,=\,\int d^2 b'\, N_{SD}\Lb Y, b'\Rb $ and
 $N_{SD}\Lb Y, b\Rb$ has been calculated in Ref.\cite{LEPP}. It has the 
form:

\bea \label{O3}
N_{SD}\Lb Y, b'\Rb \,\,&=& a \Big( a\Lb 1 - \exp\Lb - { T}\Rb\Rb\,+\,(1 - a)\frac{{ T}}{1 \,+\,{T}}\Big)^2 \,+\,a^2 (1 -  a)\frac{2 {T}^2}{\Lb 1\,+\,{ T}\Rb\, \Lb 1 \,+\,2 {T}\Rb}\nn\\
&+&\,a (1 -  a )^2\Big( \frac{{T}}{1 + { T}}\,+\,\exp\Lb 1 + \frac{1}{ T}\Rb \Gamma\Lb 0, 1 + \frac{1}{ T}\Rb\,-\,\exp\Lb \frac{1}{ T}\Rb \Gamma\Lb 0,  \frac{1}{ T}\Rb\Big)\nn\\
 &+& (1 - a)^3  \frac{1}{{ T}^2}\Big( { T}\Lb 1 + { T}\Rb \,-\,\exp\Lb \frac{1}{T}\Rb
 \Lb 1 + 2 {T}\Rb \Gamma\Lb 0,  \frac{1}{ T}\Rb\Big),
 \eea
 where $T = T\Lb Y, b \Rb$ of \eq{DP2}.  This definition of
 $T$ is valid only in the region where $T\, < \,1$. A more general
 formula is given in Ref.\cite{LEPP}. \eq{O3} 
  which sums the diagrams of 
\fig{sddd}-a, where  the double wavy lines crossed by the dashed one,
 denote the dressed Pomeron structure in terms of the produced particle.
  In Reggeon calculus it is referred to as the cut Pomeron. We would 
like to 
  emphasis that in our approach  this contribution is the 
solution
 to the equation for single diffractive production of Ref.\cite{KOLE},
 which is given in Ref.\cite{LEPP}.


The profile function for the single diffraction production
 is taken from Eq.(3.25) of Ref.\cite{GLM2}.
\subsubsection{Double diffraction}
The double diffraction  cross section has the form
\beq \label{O4}
\sigma_{DD}\Lb Y \Rb\,\,=\,\,\int d^2 b \frac{d \sigma_{DD}}{d b^2}\,\,=\,\,\int d^2\, b g^2\,S_{DD}
\Lb b \Rb \bar{N}_{DD}\Lb Y\Rb\,e^{ - 2\,\Omega\Lb Y,b\Rb}.
\eeq
$ \bar{N}_{DD}\Lb Y\Rb\,=\,\int d^2 b'\, N_{DD}\Lb Y, b'\Rb $ and $N_{DD}\Lb Y, b\Rb$ can be determined from the simple expression
\beq \label{O5}
N_{DD}\Lb Y, b'\Rb\,\,=\,\,2\,G^{\mbox{dressed}}_\pom\Lb
 T\Rb\,\,-\,\,G^{\mbox{dressed}}_\pom\Lb 2\, T\Rb,
\eeq
where  $T = T\Lb Y, b \Rb$ of \eq{DP2} with the same comments as for \eq{O3}.
\eq{O4} sums the diagrams shown in \fig{sddd}-b.
\footnote{\bf We thank our referee, for pointing out, that in our 
treatment of double diffraction, we have not included the contribution 
arising from the superposition of two (projectile and target) single 
diffraction processes, which may provide corrections to our present 
results. We will include this process in our planned two channel 
treatment.}

The profile $S_{DD}\Lb b \Rb$ is given by
\beq \label{O6}
S_{DD}\Lb b \Rb\,\,=\,\,\int d^2 b' S\Lb \vec{b}
 \,-\,\vec{b}^{\,'}\Rb\,S\Lb b' \Rb\,\,=\,\,
 \frac{m_1^5 b^3}{96 \pi} K_3\Lb m_1 b\Rb.
\eeq
\subsection{Phenomenological parameters}

In this section we summarize our phenomenological parameters and 
 provide  theoretical estimates for them.
 Altogether,  we have five parameters:
  $g$, $\phi_0$ , $\lambda$, $m$ and $m_1$. 
\begin{itemize}
\item \quad $\lambda$  in the  CGC/saturation approach, 
can be calculated in the leading order of  
perturbative
 QCD.  It characterizes the energy 
dependence
   of the saturation scale in proton-proton collisions.
 Theoretical estimates
 give $\lambda = \bas \chi\Lb \gamma_{cr}\Rb/(1 - \gamma_{cr}) \,\approx\,4.88\bas$
 in leading order of perturbative QCD, where $\gamma_{cr} = 0.37$. 
However, the
 estimates with a running QCD coupling, as well as CGC/saturation 
phenomenology,
 lead to $\lambda = 0.2 \div 0.3$.
\item \quad In the
 vicinity
 of the saturation line $G_\pom = \phi_0 \Lb r^2 Q^2_s\Rb^{(1 - 
\ga_{cr})}$ (see Eqn.{2.1} ).
$\phi_0$ denotes the value of the Pomeron Green
 function on the saturation line (at $z = 0, r Q_s = 1 $). 
 The exact value of $\phi_0$  cannot be determined without specifying the
 Pomeron-hadron interaction in more detail than we have.
 However, $\phi_0 \,\propto\,\bas^2$ so we expect  $\phi_0$ to be small.
\item \quad $m_1$ and $m$ are pure phenomenological parameters, 
 in our formulae we assumed that $m \,\gg\,m_1$. We make this assumption 
  in order to simplify the formula.
\item $g$ is a pure phenomenological parameter which
 we assumed to be larger than $G_{3 \pom}$.
\end{itemize}
\section{Fit to the data}
\subsection{Cross sections and the values of the parameters}
We determine the five parameters that our model depends on, by 
 fitting  to  experimental
 data for the following set of the observable:
total, inelastic and elastic cross sections,  for single and double
 diffractive production cross sections, and for the slope of the forward 
elastic
 differential cross section.
We fit to  the high energy data with $W \geq$ 0.546 TeV.
  We chose  the minimal energy $W = 0.546\,$ TeV  in
 our fit, as starting from this energy the CGC/saturation
 approach is able to describe the data on inclusive production
 in proton-proton collisions (see Ref.\cite{LR}). On the other
 hand the energy $W = 0.2\,$ TeV is  too low, as at this
 energy saturation occurs in ion-ion and proton-ion collisions, but
 not in proton-proton collisions \cite{KLN}.

\begin{figure}[h]
\centering
\begin{tabular}{c c c}
\includegraphics[width=0.4\textwidth, height=5.8cm]{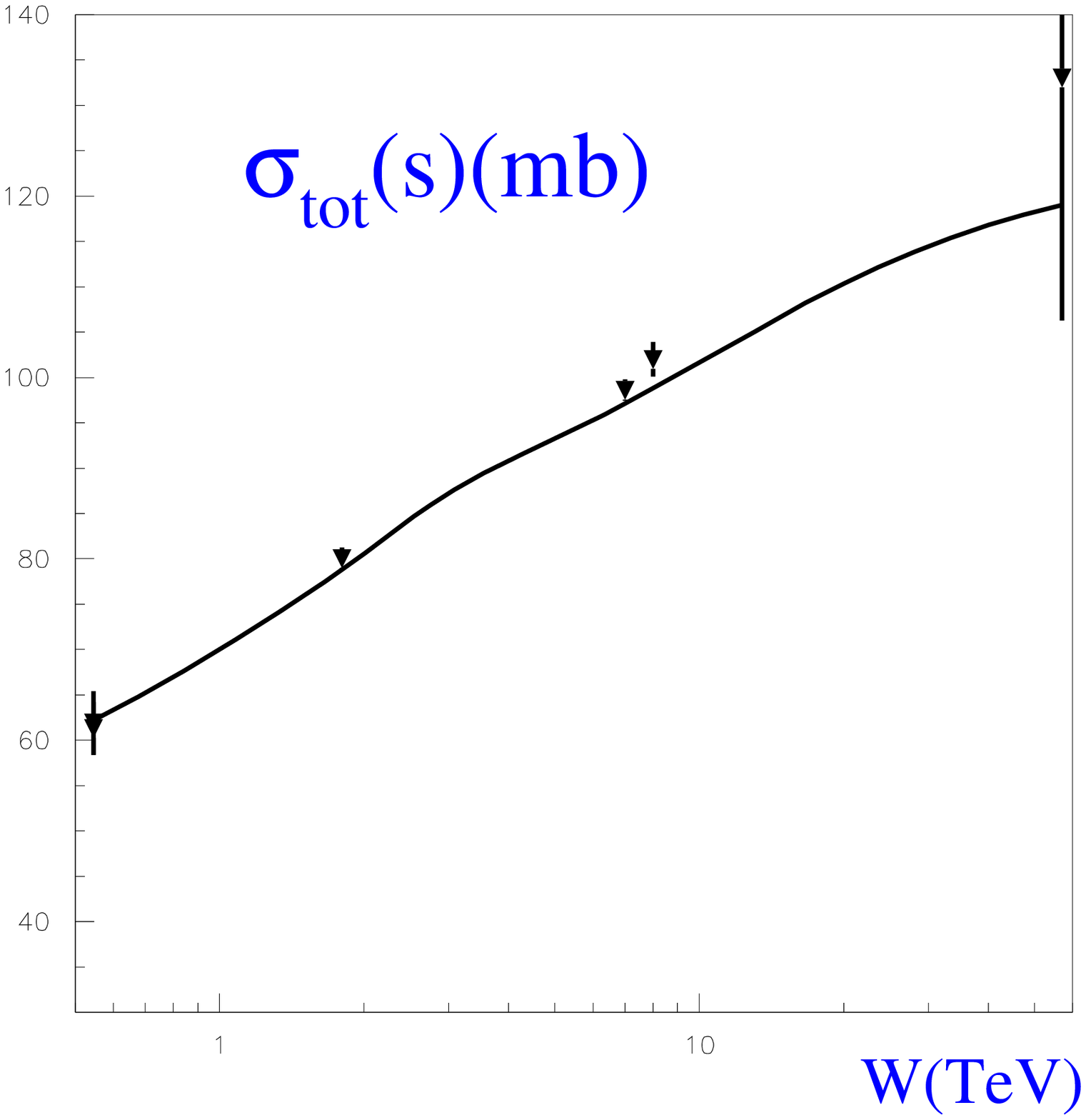}&~~~~~~~~~&
\includegraphics[width=0.4\textwidth,height=5.8cm]{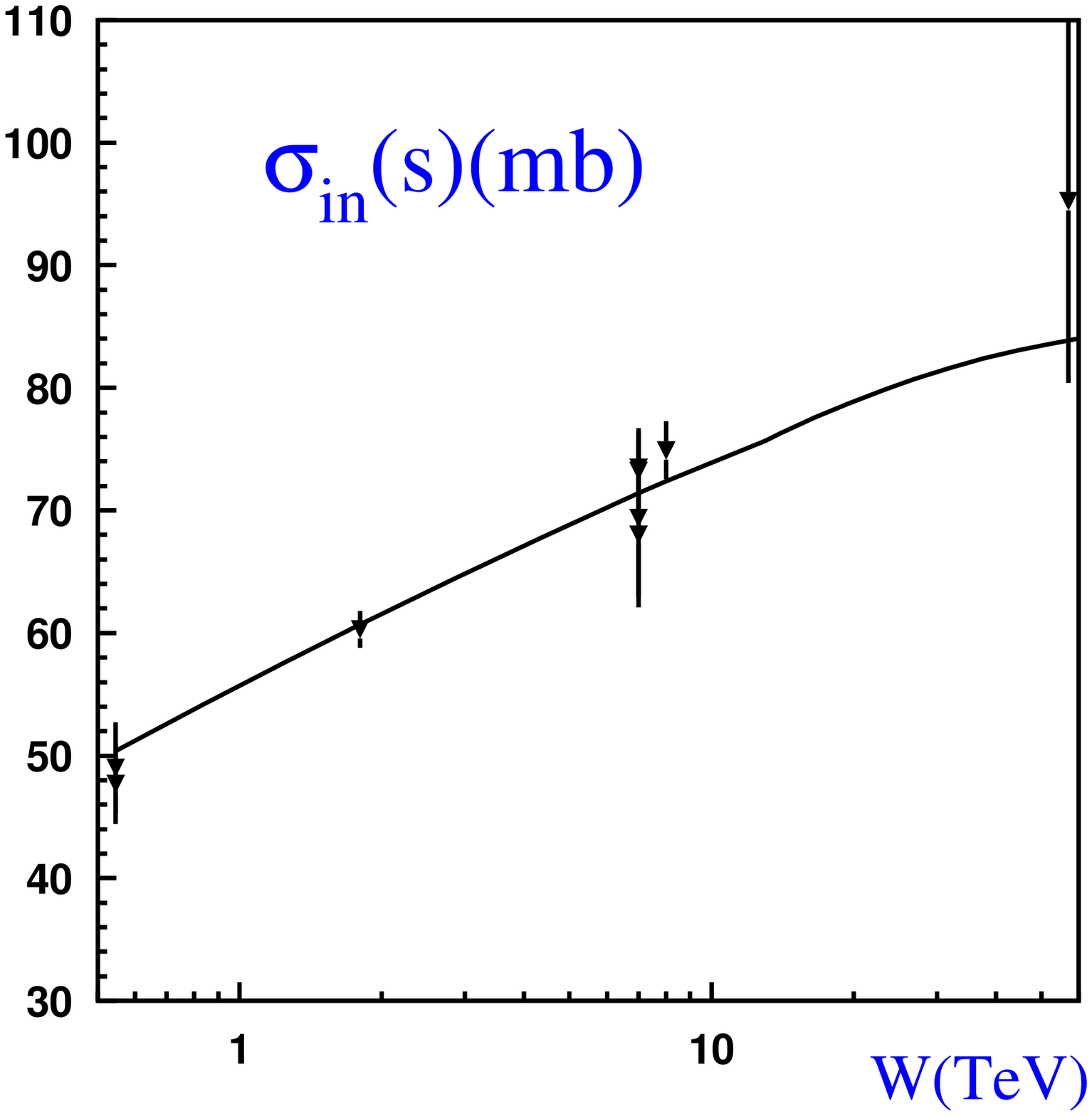}\\
\fig{fit}-a &  &\fig{fit}-b \\
\includegraphics[width=0.4\textwidth,height=5.8cm]{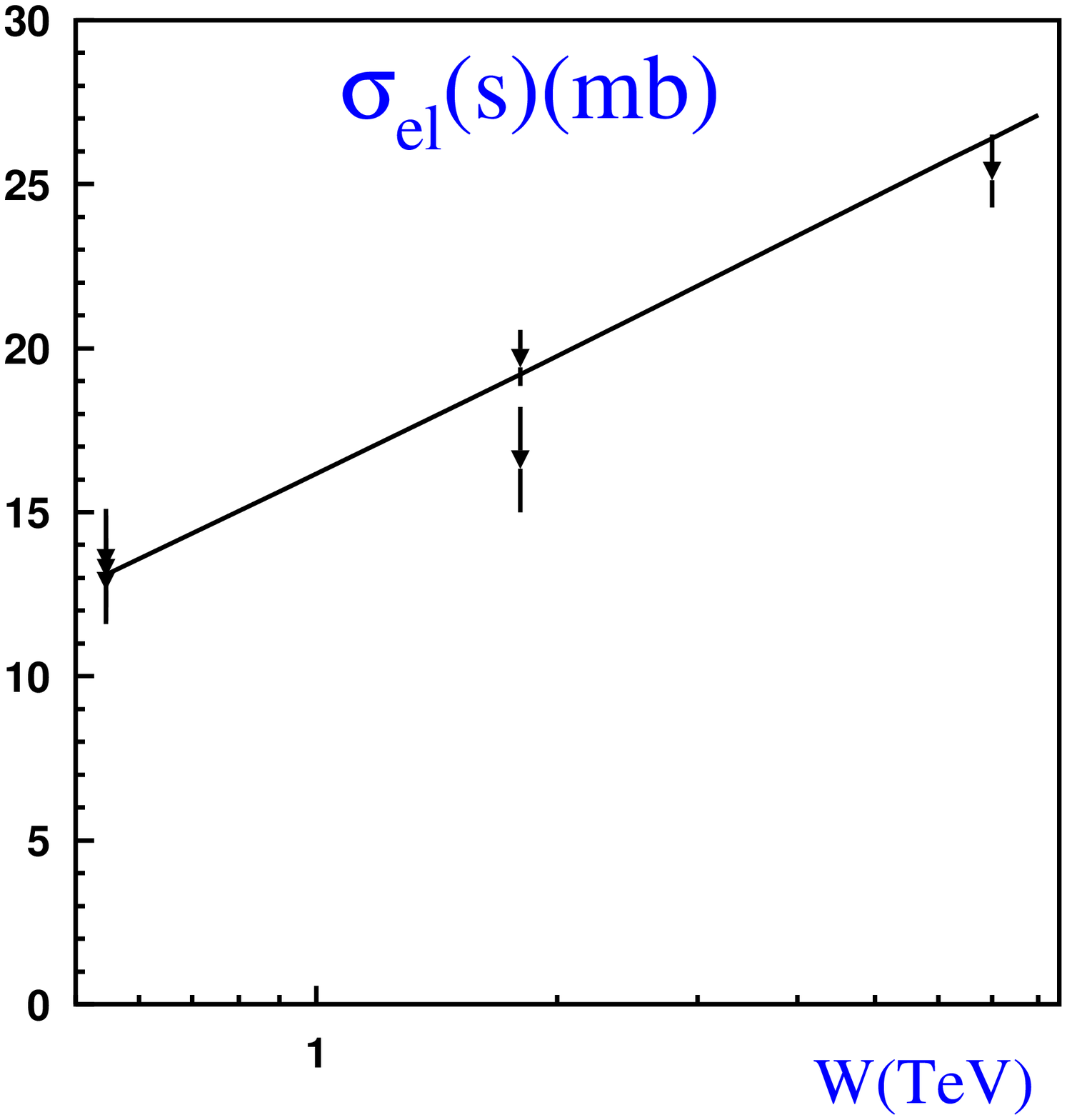}&  &
 \includegraphics[width=0.4\textwidth,height=5.8cm]{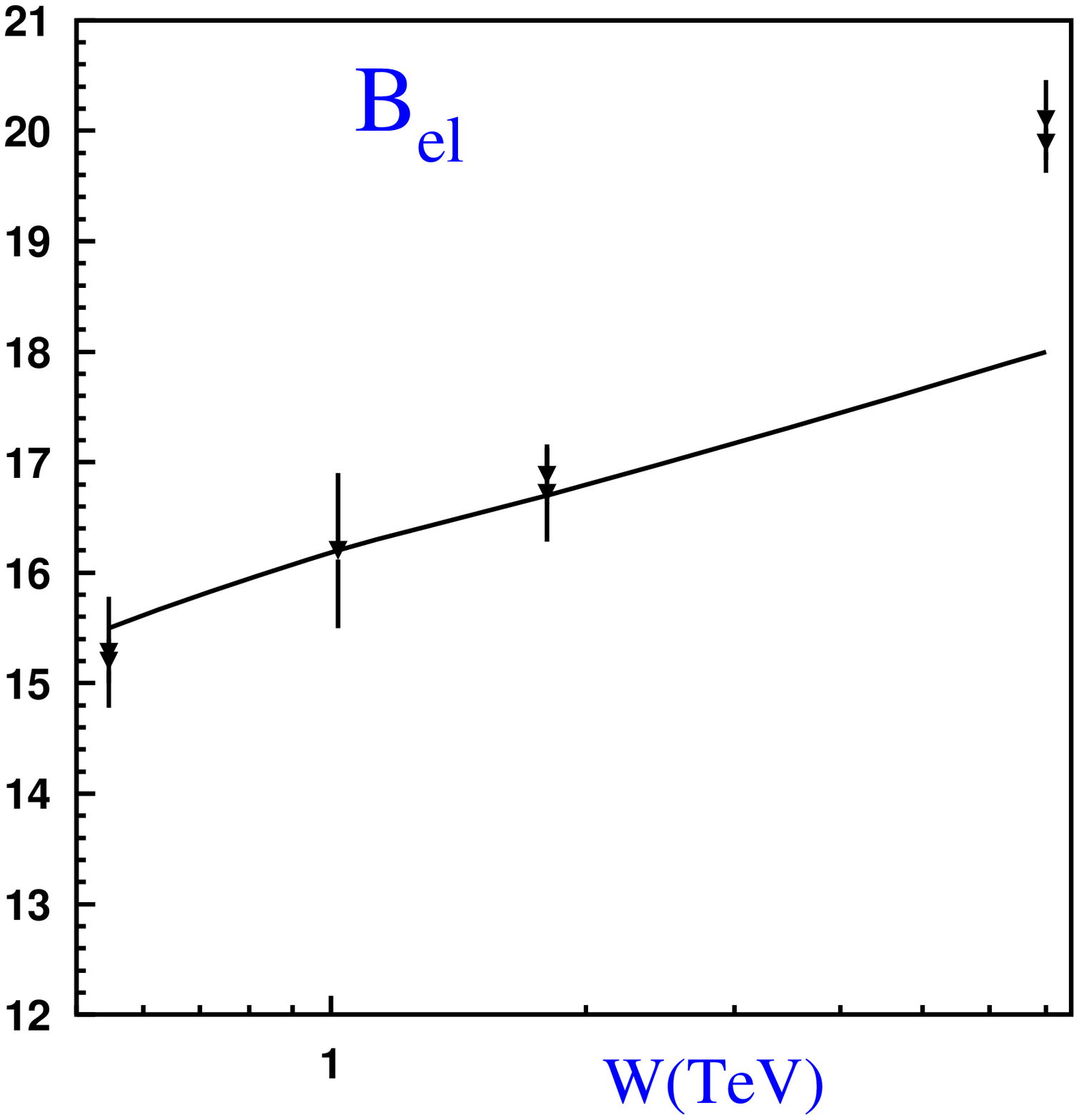}\\
 \fig{fit}-c & &\fig{fit}-d\\
 \includegraphics[width=0.4\textwidth,height=5.8cm]{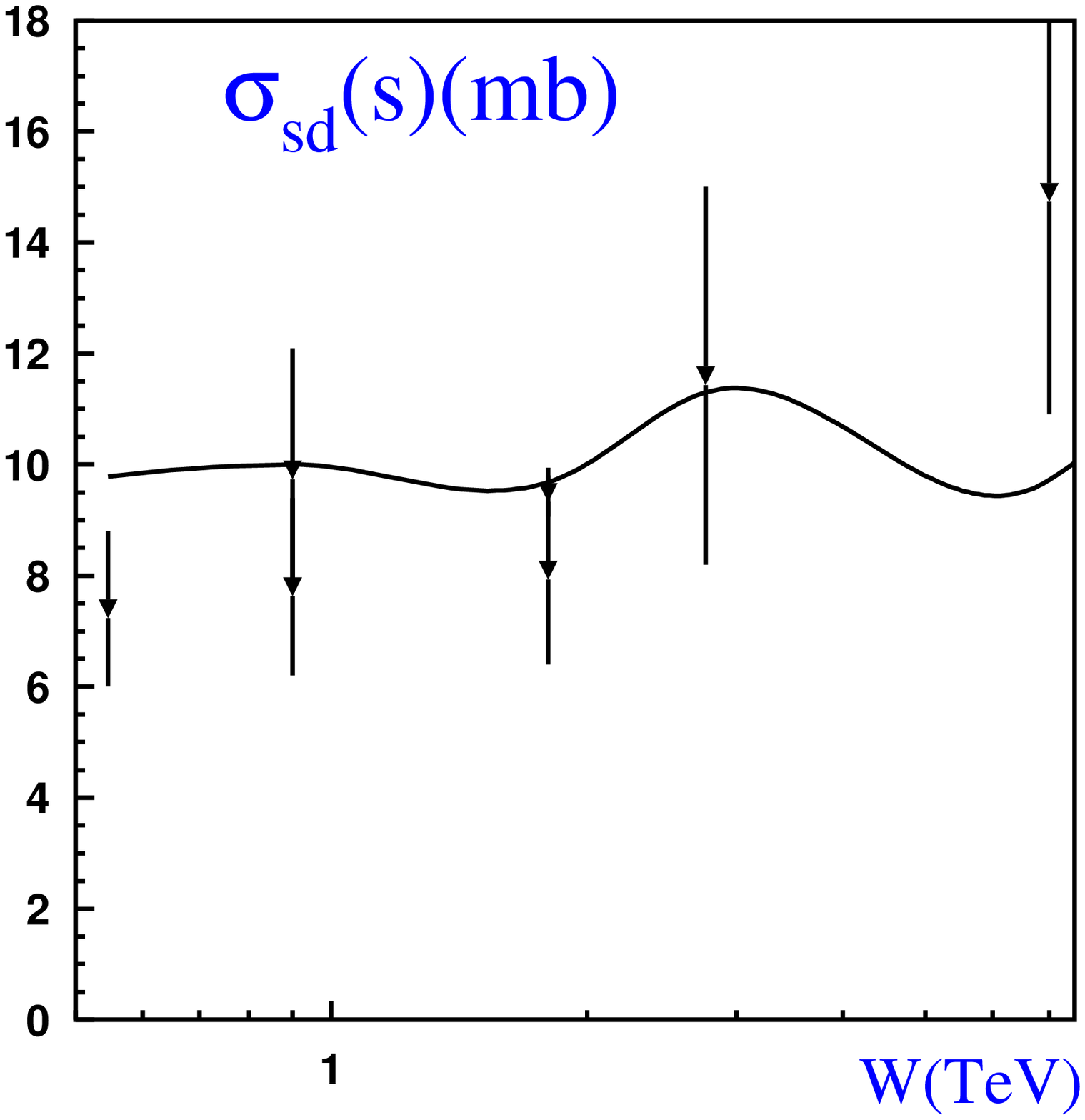}& &
\includegraphics[width=0.4\textwidth,height=5.8cm]{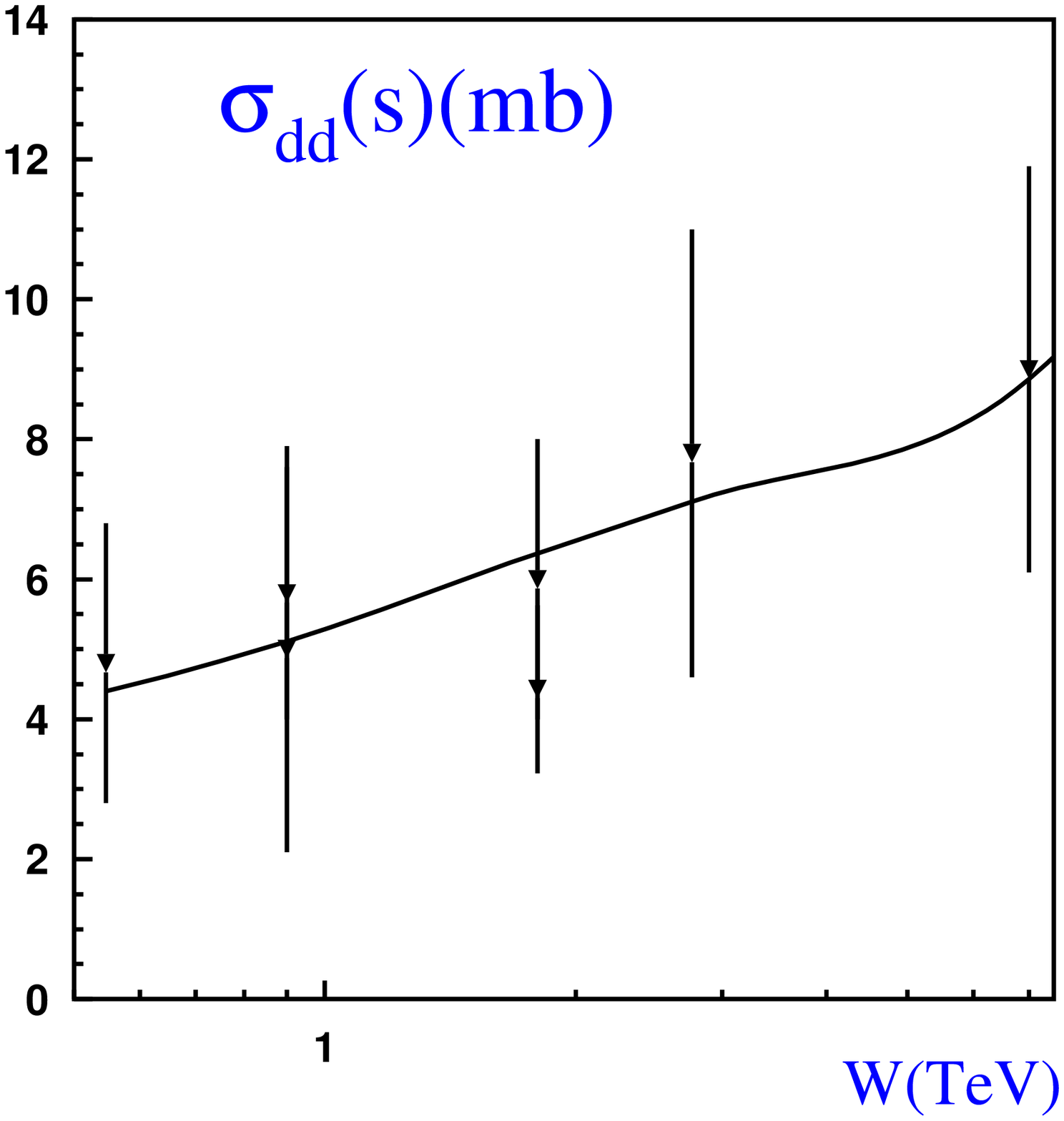}\\
\fig{fit}-e& &\fig{fit}-f \\
\end{tabular}
\caption{Comparison with the experimental data: the energy behaviour
 of the total (\protect\fig{fit}-a), inelastic  (\protect\fig{fit}-b),
 elastic cross sections   (\protect\fig{fit}-c), as well as the elastic slope ($B_{el}$,\protect\fig{fit}-d) and 
   single diffraction 
(\protect\fig{fit}-e) and 
double diffraction (\protect\fig{fit}-f) cross sections.
The solid lines show our present fit. The data has been taken from            
Ref.\protect\cite{PDG}
for energies less than the LHC energy. At the LHC energy for total and
elastic cross
section we use TOTEM data\protect\cite{TOTEM} and for single and double
 diffraction
cross sections are taken from Ref.\protect\cite{ALICE}.}
\label{fit}
\end{figure}

 The quality of the fit can
 be seen from \fig{fit} and the values of parameters are presented in 
Table 1.


\TABLE{
\begin{tabular}{|l|l|l|l|l|l|}
\hline
$\lambda $ & $\phi_0$ & g ($GeV^{-1}$)&
m (GeV) & $m_1(GeV)$ & $\chi^2/d.o.f.$ \\
\hline
0.323 &0.019&  25.7 & 6.35&0.813& 1.98\\
\hline
\end{tabular}
\caption{Fitted parameters of the model.}
\label{t1}
}
Our first observation is that the values of all parameters  are in
 agreement with our expectations given in section 3.4.  Second,
 the overall fit has $\chi^2/d.o.f. \approx   2$ of which 40\% 
is due to our failure to reproduce the TOTEM value of $B_{el}$ at W = 7 
TeV, and therefore, except for this  point , we
 have a   reasonably  good  description of the data.

We discuss some regularities in our fit,  which could be useful for 
further and more profound understanding of  the microscopic physics.

\begin{itemize}
\item \quad We obtain a  good description both of values and
 of energy dependence  for total, inelastic and elastic cross
 sections in a wide  energy range: $W = 0.546 \div 57\,TeV$.
\item \quad  At lower energies the values of $B_{el}$ are rather 
close to the
 experimental ones, but a glance  at \fig{fit}-d shows
 that the energy behavior in our model is milder than that of the 
experimental 
one.   The LHC value of $B_{el}$ is considerably
 higher than our prediction. A natural conjecture would be 
that this behaviour is a direct consequences of long standing and
 unsolved problem in the CGC/saturation approach: i.e. the large impact 
parameter
 behaviour of the BFKL Pomeron \cite{KOWI}, however, we do not think that 
this
 is correct.
 Indeed, the problem noticed in Ref. \cite{KOWI} is one of the principal
 problems of the CGC/saturation approach, and has not yet been solved. 
 Several theoretical models \cite{LLB} 
show, that the correct (exponential decreasing $~\exp(-\mu b)$) large
 impact parameter behaviour  of the scattering amplitude, does not 
influence
  the  BFKL Pomeron,     
  and  does not even produce  shrinkage of
 the diffraction peak (i.e. energy behaviour of $B_{el}$). We can also see
  indications of the weak influence of the large $b$ tail of the profile
 function in the data. Indeed, we were able to describe both $\sigma_{tot}$
 and $\sigma_{el}$ which could be possible, only if one profile function
 describes  the typical $b$ behaviour correctly. Since there is no 
reason to
 expect that large $b$ behaviour could be crucial in the description of 
the
 available data, we assume that the CGC/saturation approach will not be 
changed by the (incorrect) large $b$ behaviour of the BFKL Pomeron, and 
that all 
non-perturbative corrections have to be included in $Q_s$ \cite{IIM}:  
 the only dimensional parameter of the theory. It should be stressed
 that this assumption leads to correct exponential fall off of the
 amplitude in the framework of our approach.  We believe that the weak
energy
 dependence of the elastic slope,  follows mostly from rather
 small shadowing corrections (see \fig{ael}-b) which we hope will be 
alleviated
 in our future planned two channel treatment.
\item \quad  Our single diffractive production cross section  has
 values within the experimental error, but our model 
 predicts a lower cross section at $W =$  7 TeV, than the one given by
 ALICE\cite{ALICE}. The model's results are shown in Table 2
and  support the idea that the single diffraction is saturated at
 high energy\footnote{ As far as we know, K. Goulianos was the
 first to predict this feature of   single diffraction
 production at high energies \cite{DINODIF}, based on  a different
 point of view of high energy interactions, than ours.}.   In our model
 the  saturation, and oscillatory behaviour, of the  single diffraction 
cross section stems from two
 sources: the simple one channel parametrization and the robust features 
of
 CGC/saturation approach. 
 We plan to develop a more sophisticated (say two channel) model to
 disentangle these two effects, and we trust that this will also cure the 
present oscillatory 
behaviour.
\item \quad The striking feature of our model and, perhaps of the data, is
 that the double diffractive cross section increases with energy (see Table 2).
\end{itemize}

\TABLE{
\begin{tabular}{|l|l|l|l|l|l|}
\hline\hline
$W(TeV) $ & $\sigma_{tot}$(mb) & $ \sigma_{el}$(mb)& $B_{el}(GeV^{-2})$&
$\sigma_{sd}(mb)$ & $\sigma_{dd}(mb)$ \\
\hline
0.546&  62.2 &13.1& 15.5 & 9.78 & 4.4\\
\hline
0.9 & 69.1 & 16 & 16 &10 & 5.11\\
\hline
1.8 & 78.8 & 19.2 & 16.7 &9.69 & 6.37\\
\hline
2.76 & 84.6 &21.4 &17.1 & 10.9 & 7.09\\
\hline
7 & 97.1 & 26.4 & 18 & 9.71 &  8.86\\
\hline
8 & 98.8 &  27.1 & 18.2 & 10.8 &  8.86\\
\hline
13 &105  & 29.5  &18.6  &  9.7 & 9.81\\
\hline
14 & 106 & 29.9 &18.7 & 9.5 &  9.98\\
\hline
57& 120 & 35.7 &19.9 &10 & 13\\
\hline
\hline
\end{tabular}
\caption{ Cross sections at high energies predicted by the model.}
\label{t2}
}
\subsection{Partial amplitudes and comparison with other models}
\subsubsection{Partial amplitudes}
\begin{figure}[ht]
    \centering
    \begin{tabular}{c c}
  \leavevmode
      \includegraphics[width=8cm]{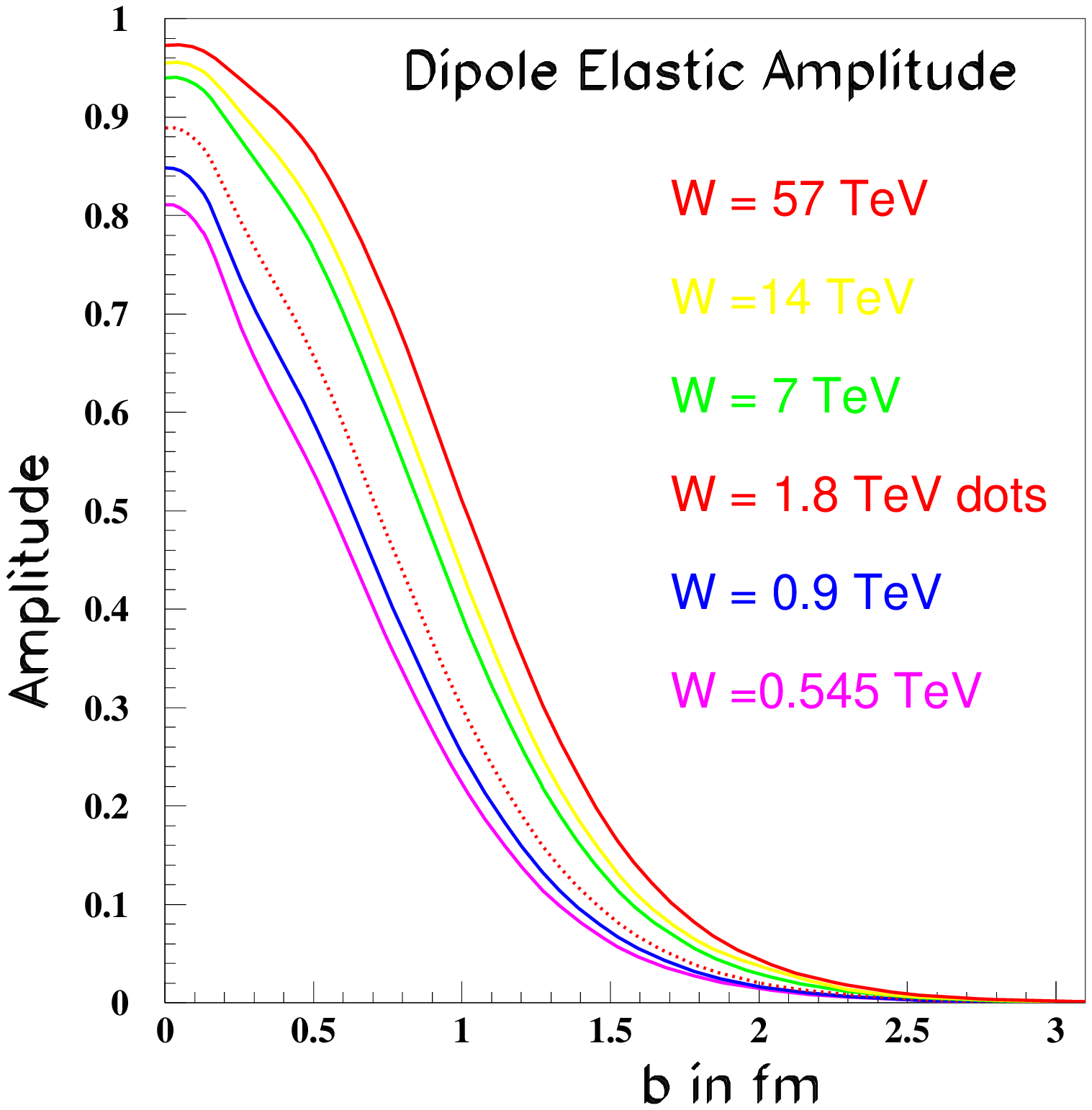}&  \includegraphics[width=8cm,height=7cm]{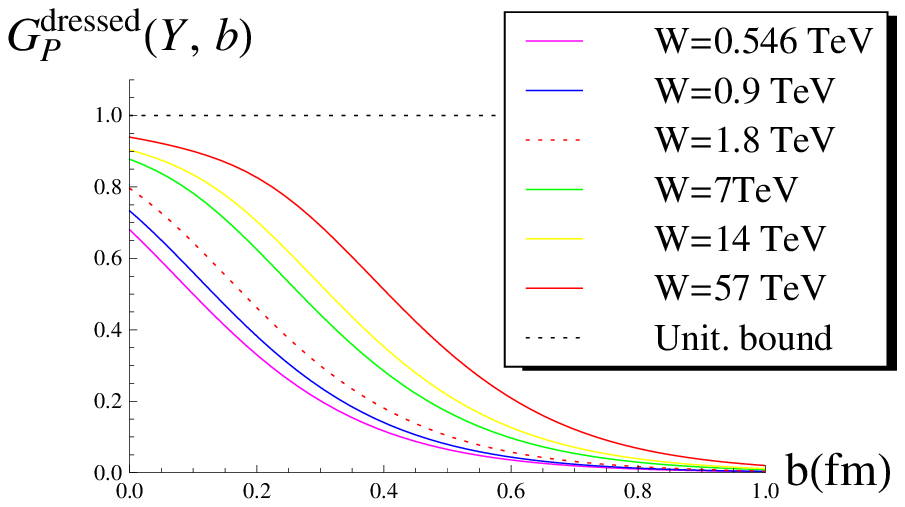} \\
      \fig{ael}-a & \fig{ael}-b\\
      \end{tabular}
           \caption{The behaviour of the elastic amplitude $A_{el}\Lb s, b\Rb$ versus $b$(\protect\fig{ael}-a) and the $b$-dependence of $G^{\mbox{dressed}}_\pom\Lb Y, b\Rb$ at different energies.}
\label{ael}
   \end{figure}

. 
We believe that the information contained in the  impact parameter 
dependence
 of partial amplitude, is  instructive for understanding the nature of 
 strong interactions at high energy. It is also useful for
 illustrating  the strong and weak  features
 of the model. Our present model, which is a single 
channel model has only 
 one elastic 
 amplitude, $A_{el}\Lb s, b\Rb$  which is shown in \fig{ael}-a.

 
 One can consider our  proton as a gray disk ($ A_{el}\Lb s, b =0\Rb 
\,<\,1$), 
even at energies as high 
 as $W = 57\,TeV$. This  behaviour   stems mostly   from the 
$b$-dependence of the Green function of the dressed Pomeron (see 
\fig{ael}-b).
 
 The typical $b$ increases with energy. Note, that such an increase has 
been
 seen in the behaviour of $B_{el}$ versus energy (see \fig{fit}-d). This
 behaviour is due to  strong saturation in  dipole-dipole 
scattering,
 as the slope of the Pomeron trajectory  for the BFKL Pomeron, is equal to 
zero.
     \begin{figure}[ht]
    \centering
    \begin{tabular}{c c}
  \leavevmode
      \includegraphics[width=8cm]{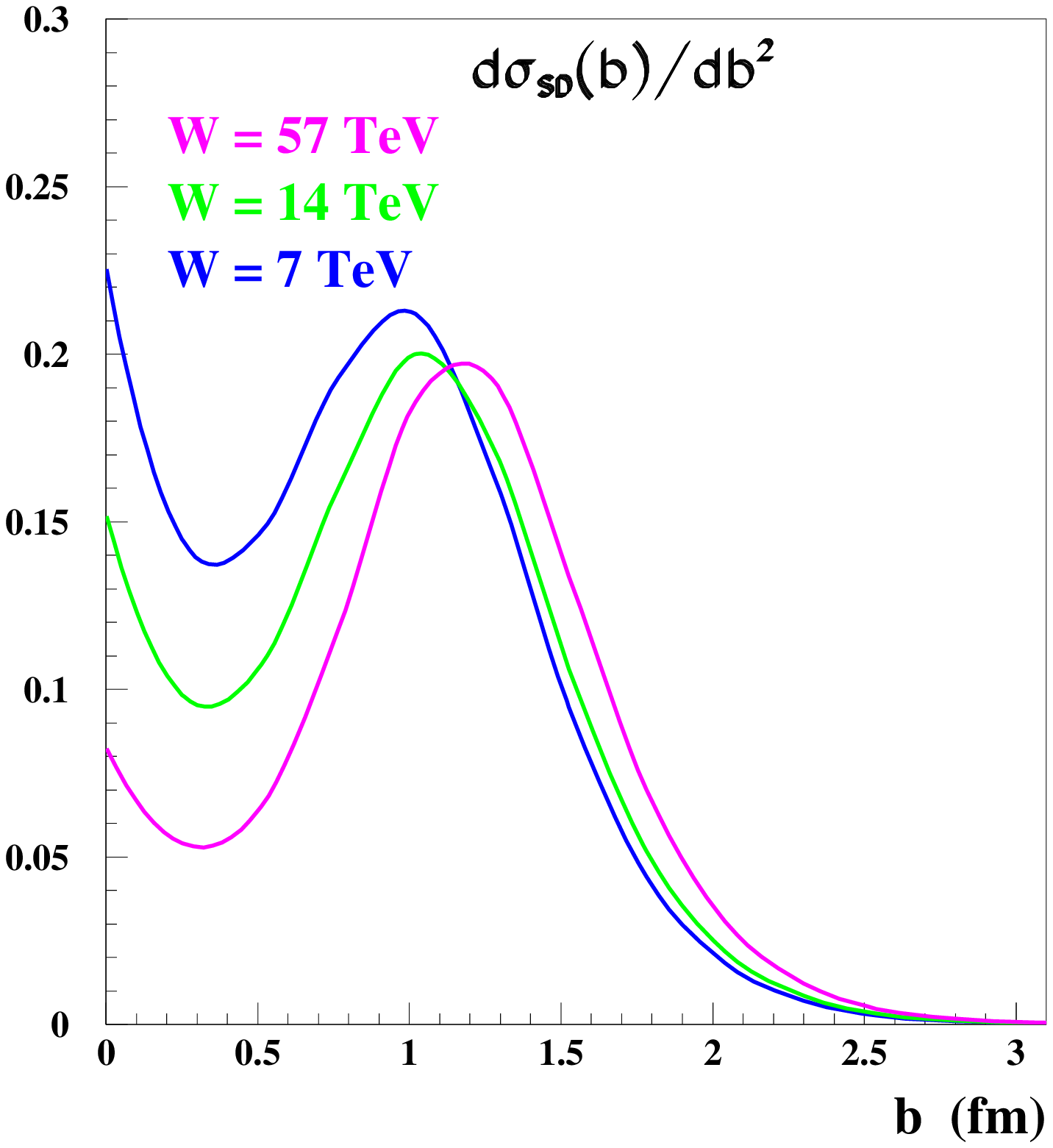}  &      
\includegraphics[width=8cm]{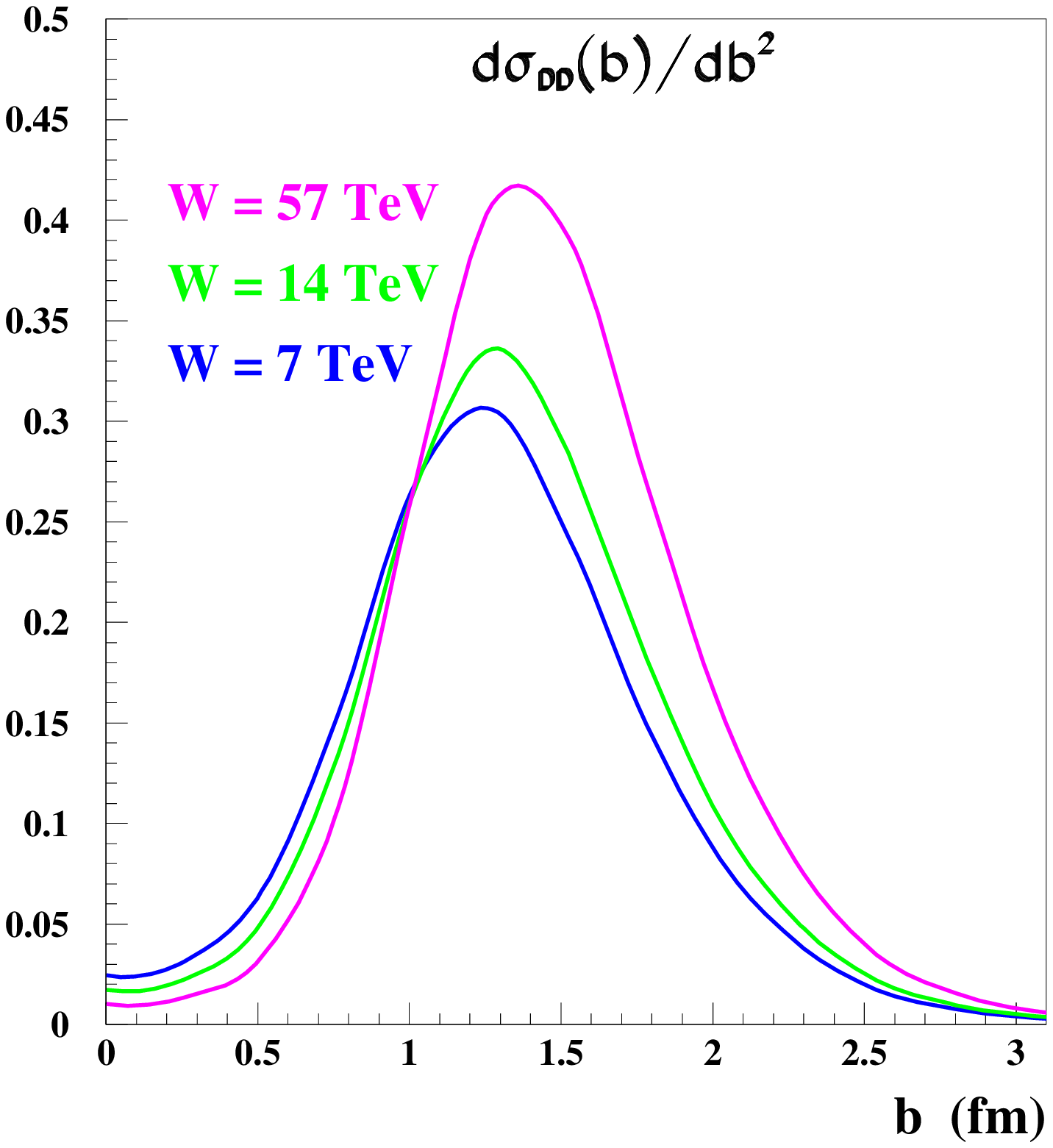} \\
      \fig{asddd}-a &\fig{asddd}-b\\
      \end{tabular}
            \caption{ $d \sigma_{SD}\Lb s, b\Rb/d b^2$ (see \protect\eq{O2}, \protect\fig{asddd}-a) and $d \sigma_{DD}\Lb s, b\Rb/d b^2$ (see \protect\eq{O4}, \protect\fig{asddd}-b)  versus $b$ at different energies}
\label{asddd}
   \end{figure}

 
 In \fig{asddd}  $d \sigma_{SD}\Lb s, b\Rb/d b^2$(\eq{O2}) and
 $d \sigma_{DD}\Lb s, b\Rb/d b^2$ (\eq{O4}) 
 are plotted. Note that the  single and double diffraction
 production have quite different distributions in $b$.
 $d \sigma_{DD}\Lb s, b\Rb/d b^2$,  
 as  is expected,  has a peripheral
 form having a minimum at $b =0$, 
 the maximum and the width  of  the $b$-distribution of
 $d \sigma_{DD}\Lb s, b\Rb/d b^2$ grows considerably 
 with energy. On the other hand, the peripheral nature of
 the single diffractive production starts to appear only
 at high energies. The typical distribution has two maxima
 at $b=0$ and $b \approx 1 \,fm$, both  decrease with
 energy, while the width of the distribution slowly 
  increases 
 with  energy.  Such unexpected behaviour stems mostly  from
 the rather transparent dipole-dipole interaction, which is
 due to the values of the  fitted parameters in our model.
 
  The feedback from  the impact parameter
 behaviour of the scattering amplitude is that there is a  need to 
increase 
the shadowing
 corrections in the  dipole-dipole scattering. From our experience with 
the description
 of soft interactions at high energies, we know that one way of achieving 
this, is by using 
  a two channel model.

  \subsubsection{Comparison with other models on the market}
\begin{figure}[ht]
    \centering
    \begin{tabular}{c c}
  \leavevmode
      \includegraphics[width=7cm]{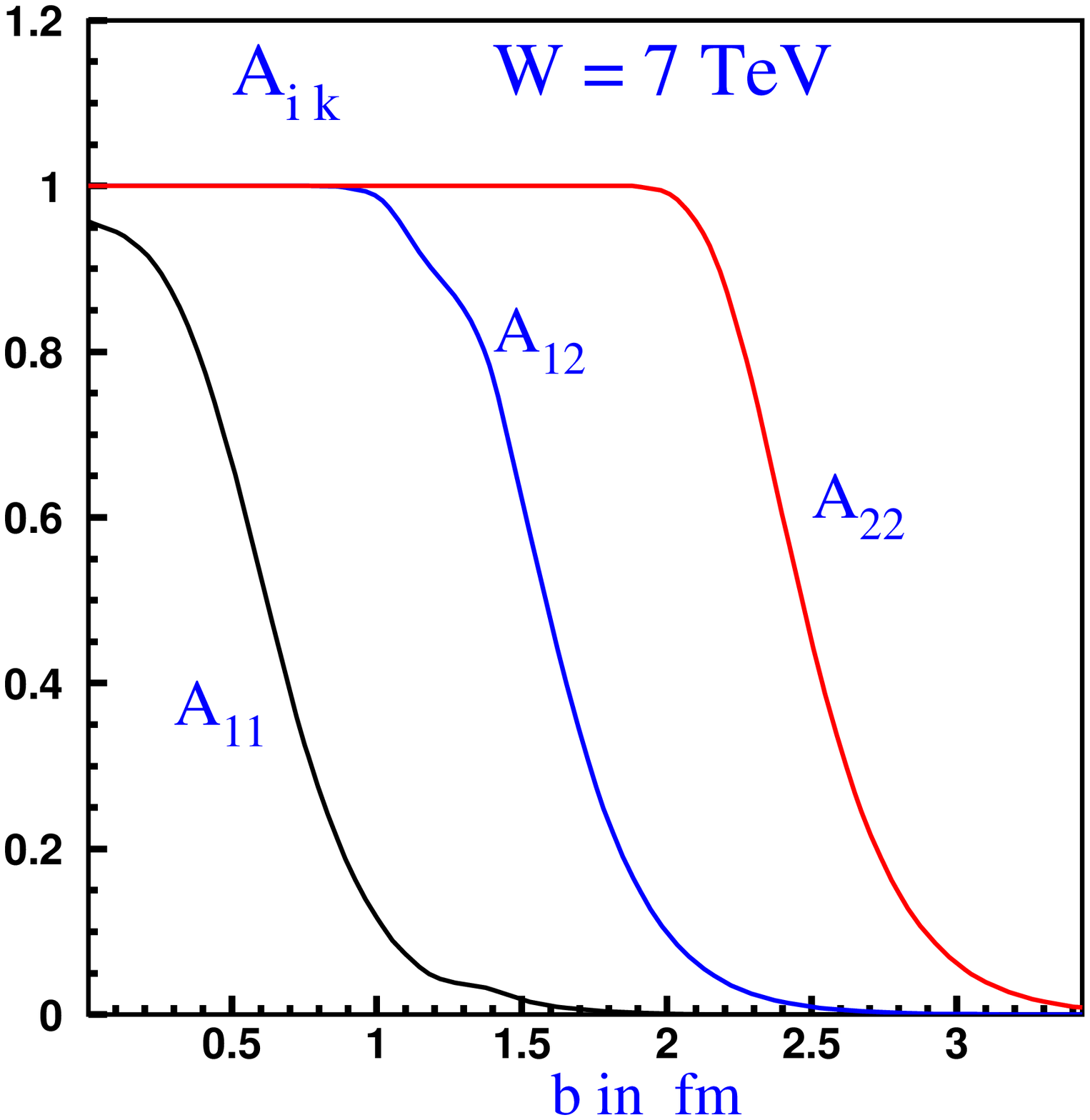}  &      \includegraphics[width=7.25cm]{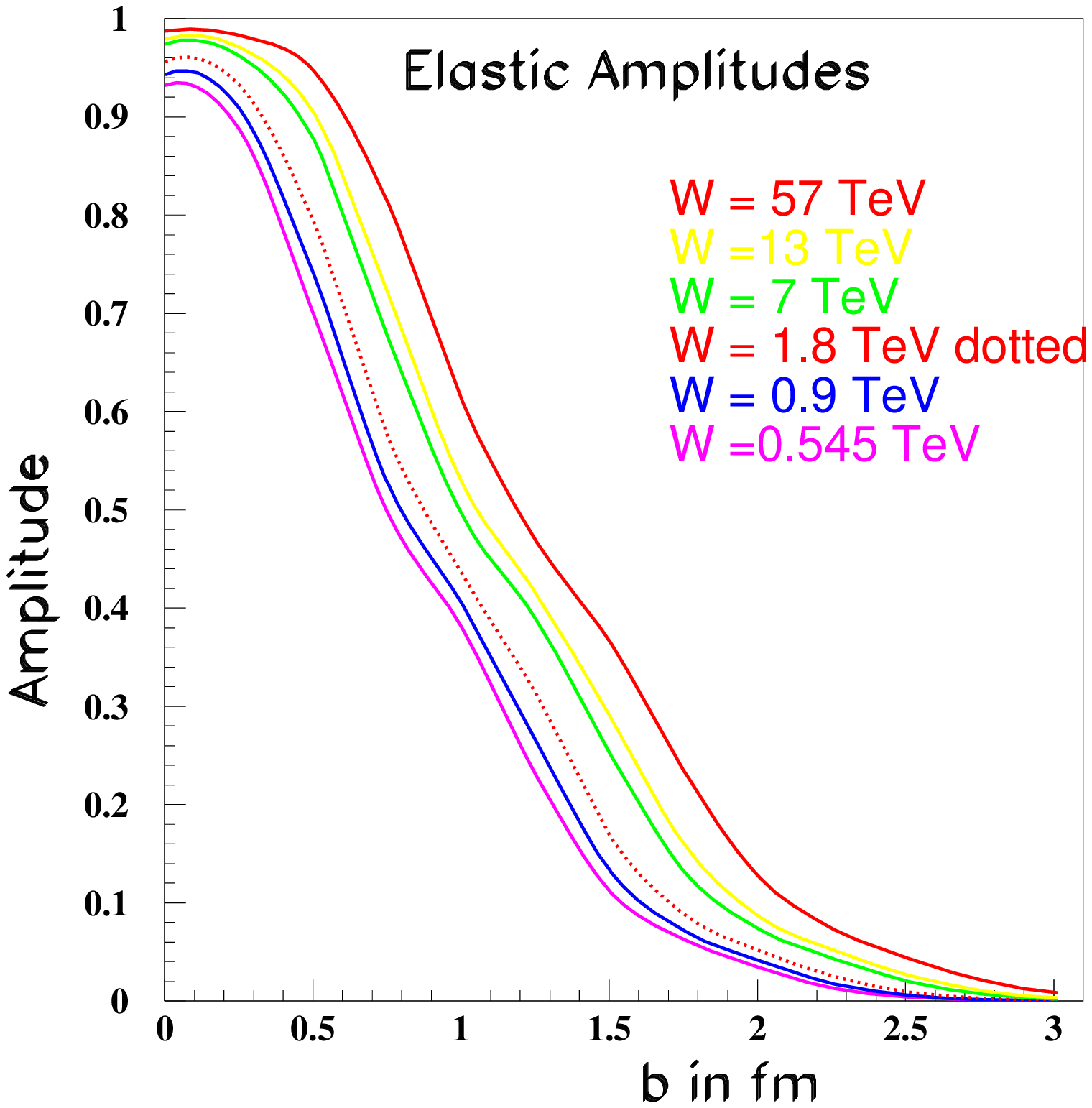} \\
      \fig{glm}-a &\fig{glm}-b\\
      \end{tabular}
            \caption{ Partial amplitudes of the GLM two channel model of 
Ref.\cite{GLMREV}. }
\label{glm}
   \end{figure}

\begin{figure}[ht]
    \centering
  \leavevmode
      \includegraphics[width=10cm]{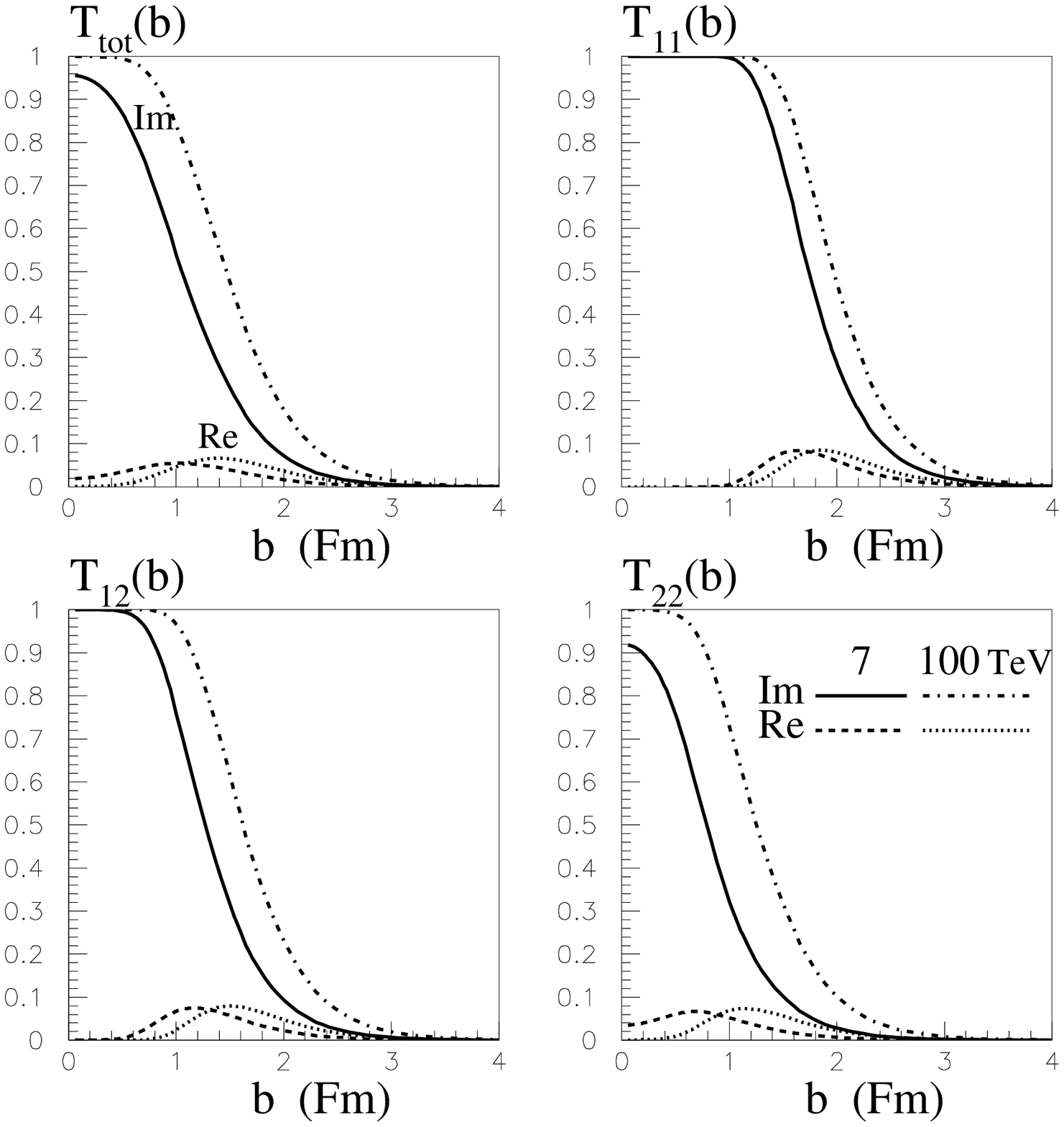}   
                 \caption{ Partial amplitudes  ($T_{ik}\Lb b \Rb \equiv 
A_{ik}\Lb b \Rb$)  of  the Durham group's  model \protect\cite{KMRREC}.
 The figure is taken from Ref.\protect\cite{KMRREC}}
\label{kmr}
   \end{figure}

  In brief, this model is a one channel eikonal - type model,
 with a dressed Pomeron 
whose form has been adapted from  the CGC/saturation approach.
  It differs from  
 our previous model 
(see review \cite{GLMREV}) which is a two component model having
 three different 
partial amplitudes (see \fig{glm}-a).
  The striking feature of the two component model is that two
 amplitudes  become 
black  at 
low energies. The resulting elastic amplitude in the two component
  model is shown in 
\fig{glm}-b, and is similar to that of
  our present model. This is not surprising, as both models provide
 a good description 
of the data.
  However, the fact that in our present approach none of ingredients:
 elastic amplitude 
and Green's function of dressed Pomeron, reach the black limit,
 looks surprising, especially so, since the recent model proposed by KMR
  (see Ref.\cite{KMRREC} and \fig{kmr}), supports the fact that two of the
 partial amplitudes are black.

    We wish to emphasis,
 that  our two models: the present one and the two component model,
 lead to completely
 different predictions for single diffraction: in the first the cross section
 is saturated, while in the second it grows with energy.
  
Over the past few years a number of models have been constructed
\cite{OST,KMRREC,KAPO} based on Reggeon Field Theory whose results 
for energies below that of the LHC are similar  i.e.
 they adhere to the general trend of 
the experimental data, in that their results for $\sigma_{tot},
 \sigma_{el}, B_{el},
\sigma_{SD}$ and $\sigma_{DD}$ increase with increasing energy.
 The same applies to the 
Monte Carlo programs MBR\cite{DINO} (which is 
an "event generator" based on an enhanced PYTHIA8 simulation) and QGSJET-II 
\cite{OSTREC}.

\par Following the appearance of the preliminary results for single and double 
diffraction cross sections by the TOTEM Collaboration\cite{TOTEMREC}
 and the CMS Collaboration \cite{CMSREC} 
which suggests that the growth of $\sigma_{SD}$ and $\sigma_{DD}$ maybe
 leveling off 
(or even decreasing) at W = 7 TeV. KMR \cite{NEWKMR} have modified 
their model by including energy dependent couplings, so as to be in accord
 with the 
TOTEM results. We would like to stress that the 
published results of the ALICE collaboration \cite{ALICEREC} have the 
single and double diffractive cross  sections still increasing at LHC energies.

       
     \section{Conclusions} 
 In this paper we present a first attempt to develop a consistent approach 
based on the BFKL Pomeron and the CGC/saturation approach for  soft
 interactions at high energy.   We follow our
 general strategy for constructing models for strong interactions at
 high energy i.e. to maximize the theoretical ingredients,
 and to minimize the number of  phenomenological
 parameters. 
 
  We construct an  eikonal-type model whose opacity  is determined
 by the exchange of the dressed BFKL Pomeron. The  Green's function of
 the Pomeron is calculated in the framework of the CGC/saturation 
approach.
 Having only five parameters we obtain  a reasonable description of the 
experimental
 data at high energies ( $W\,\geq\,0.546\,TeV$).
 One of these five parameters $\lambda$, determines the energy
 dependence of the saturation scale, its value $\lambda = 0.323$
 is a bit higher than the values that has been found from the description
 of the DIS and heavy ion scattering data, but it is close to them.
 
 Using the value of $\lambda$ from the fit we can estimate the value of the
 intercept of the BFKL Pomeron since $\lambda = 4.88 \,\bas$ while
 $\Delta_{\mbox{\tiny BFKL}} = 2.8\,\bas \,\approx\,0.2$. From \eq{MPSI2}
 we see that we can trust the MPSI approximation for 
 $Y\,\,\leq\,\,$ 36,
 and therefore,  the MPSI approximation provides the exact answer
 for the entire kinematic region
 of energies quoted.
    
    In our model we find different behaviour for the
   single and double diffraction cross sections at high energies.
 The single diffraction reaches a saturated value (about 10 mb) at
 high energies, while the double diffraction cross section  grows 
steadily. The reason for this, is the different  energy and 
impact parameter dependences, of the  diagrams describing $\sigma_{sd}$
(Fig.4-a) and $\sigma_{dd}$ (Fig.4-b). 
   
   It turns out that in the model, all ingredients are far  from
 being a black disc, in contradiction to our previous model. This
  illustrates how  important it is to find a theoretical
 approach for  soft processes.
   
   We consider this paper as a  first attempt to expand
 the CGC/saturation approach to  describe  soft
 processes at high energy. We plan to include more details
 of the CGC/saturation theory in our model and, in particular,
 to account for the running QCD coupling and  to develop a
 two channel model to disentangle the two effects: the simple eikonal
 approach and the CGC/saturation features, which are included
 in our model. This
 paper provides an  illustration of how the LHC data has stimulated 
our 
thinking.
  
  \section{Acknowledgements}
   We thank our colleagues at Tel Aviv university and UTFSM for
 encouraging discussions. Our special thanks go to Carlos Contreras,
 Alex Kovner and Misha Lublinsky for elucidating discussions on the
 subject of this paper.
   This research was supported by the BSF grant   2012124  and the
 Fondecyt (Chile) grant 1140842.


\begin{thebibliography}{99}

\bibitem{COL}
P.D.B. Collins, {\it "An introduction to Regge theory and high energy physics"}, 
Cambridge University Press 1977.
\bibitem{SOFT}
Luca Caneschi (editor), {\it "Regge Theory of Low -$p_T$ Hadronic Interaction"}, 
North-Holland 1989.
\bibitem{LEREG}
E. Levin,
{\it "An introduction to Pomerons"}, arXiv:hep-ph/9808486; 
{\it "Everything about Reggeons. I: Reggeons in *soft* interaction"}, 
arXiv:hep-ph/9710546.


\bibitem{ALICE}
M.~G.~Poghosyan,
  J.\ Phys.\ G G {\bf 38}, 124044 (2011)
  [arXiv:1109.4510 [hep-ex]].
 ALICE ~Collaboration,
  {\it ``First proton--proton collisions at the LHC as observed with the ALICE
  detector: measurement of the charged particle pseudorapidity density at
  $\sqrt{s}$ = 900 GeV,''}
  arXiv:0911.5430 [hep-ex].
\bibitem{ATLAS}
G.~Aad {\it et al.}  [ATLAS Collaboration],
  Nature Commun.\  {\bf 2},  463 (2011)
  [arXiv:1104.0326 [hep-ex]].
\bibitem{CMS}
CMS Physics Analysis Summary:
``Measurement of the inelastic pp cross section at √s = 7 TeV with the CMS detector", 2011/08/27.


\bibitem{TOTEM}
 F.~Ferro [TOTEM Collaboration],
  AIP Conf.\ Proc.\  {\bf 1350}, 172 (2011) ;\,\,\,G.~Antchev {\it 
et al.}  [TOTEM Collaboration],
  Europhys.\ Lett.\  {\bf 96},  21002 (2011),
  {\bf 95},  41001 (2011)
  [arXiv:1110.1385 [hep-ex]].


\bibitem{DL}
A. Donnachie and P.V. Landshoff,
Nucl. Phys. {\bf B231}, (1984) 189; 
Phys. Lett. {\bf B296}, (1992) 227; 
Zeit. Phys. {\bf C61}, (1994) 139.

\bibitem{GLM1}
 E.~Gotsman, E.~Levin and U.~Maor,
  Eur.\ Phys.\ J.\  C {\bf 71},  1553 (2011)
  [arXiv:1010.5323 [hep-ph]].
\bibitem{GLM2}
   E.~Gotsman, E.~Levin, U.~Maor and J.~S.~Miller,
  Eur.\ Phys.\ J.\  C {\bf 57},  689 (2008)
  [arXiv:0805.2799 [hep-ph]].
\bibitem{KAP}
 A.~B.~Kaidalov and M.~G.~Poghosyan,
  arXiv:0909.5156 [hep-ph].

\bibitem{KMR}
 A.~D.~Martin, M.~G.~Ryskin and V.~A.~Khoze,
  arXiv:1110.1973 [hep-ph].
\bibitem{OST}
S.~Ostapchenko,
Phys.Rev.\,D {\bf 81}, 11402 (2010).

\bibitem{NEWGLM}
E.~Gotsman, E.~Levin and U.~Maor,
  Phys.\ Lett.\ B {\bf 716} (2012) 425
  [arXiv:1208.0898 [hep-ph]].

\bibitem{NEWKMR}
V.~A.~Khoze, A.~D.~Martin and M.~G.~Ryskin,
  Eur.\ Phys.\ J.\ C {\bf 74} (2014) 2756
  [arXiv:1312.3851 [hep-ph]].

\bibitem{OSTREC}
S. Ostapchenko, Phys.\ Rev.\  D {\bf 83},  014018  (2011)
  [arXiv:1010.1869 [hep-ph]].

\bibitem{GLMREV}
E.~Gotsman, E.~Levin and U.~Maor,
  {\it ``A comprehensive model of soft interactions in the LHC era,''}
  arXiv:1403.4531 [hep-ph].


\bibitem{BST}
 R.~C.~Brower, J.~Polchinski, M.~J.~Strassler and C.~I.~Tan,
  JHEP {\bf 0712} (2007) 005
  [arXiv:hep-th/0603115];\,\,
R. C. Brower, M. J. Strassler and C. I. Tan,
  JHEP {\bf 0903 } (2009)  092
  [arXiv:0710.4378 [hep-th]];\,\,\, 
  JHEP {\bf 0903 } (2009)  050.
  [arXiv:0707.2408 [hep-th]].

\bibitem{POST}
 J.~Polchinski and M.~J.~Strassler,
  JHEP {\bf 0305} (2003) 012
  [arXiv:hep-th/0209211];\,\,
  Phys.\ Rev.\ Lett.\  {\bf 88} (2002) 031601
  [arXiv:hep-th/0109174].
\bibitem{BFKL4}
 A.~V.~Kotikov, L.~N.~Lipatov, A.~I.~Onishchenko and V.~N.~Velizhanin,
  Phys.\ Lett.\  B {\bf 595} (2004) 521
  [Erratum-ibid.\  B {\bf 632} (2006) 754]
  [arXiv:hep-th/0404092];\,\,
A.~V.~Kotikov and L.~N.~Lipatov,
  Nucl.\ Phys.\  B {\bf 661} (2003) 19
  [Erratum-ibid.\  B {\bf 685} (2004) 405]
  [arXiv:hep-ph/0208220];\,\,
  A.~V.~Kotikov and L.~N.~Lipatov,
  Nucl.\ Phys.\  B {\bf 582} (2000) 19
  [arXiv:hep-ph/0004008].
\bibitem{BFKL}
 E. A. Kuraev, L. N. Lipatov, and F. S. Fadin, {\it  Sov. Phys.
JETP}
                {\bf 45}, 199 (1977); \,\,\,
Ya. Ya. Balitsky and L. N. Lipatov,
               {\it   Sov. J. Nucl. Phys.}\, {\bf 28}, 22 (1978).

\bibitem{LI} 
L. N. Lipatov,
Phys. Rep. {\bf 286} (1997) 131; Sov. Phys. JETP {\bf 63} (1986) 904 
and references therein. 
\bibitem{BART}
J. Bartels, M. Braun and G. P. Vacca,
Eur. Phys. J. {\bf C40} (2005) 419 [arXiv:hep-ph/0412218].
J. Bartels and C. Ewerz,
JHEP {\bf 9909} 026 (1999) [arXiv:hep-ph/9908454]. 
J. Bartels and M. Wusthoff,
Z. Phys. {\bf C6}, (1995) 157. 
A. H. Mueller and B. Patel,
Nucl. Phys. {\bf B425} (1994) 471 [arXiv:hep-ph/9403256].
J. Bartels,
Z. Phys. {\bf C60} (1993) 471. 

\bibitem{GLMONEPOMERON}
E.~Gotsman, E.~Levin and U.~Maor,
  {\it ``A Soft Interaction Model at Ultra High Energies: Amplitudes, Cross Sections and Survival Probabilities,''}
  arXiv:0708.1506 [hep-ph].

\bibitem{FROI}
M.~Froissart,
{\it Phys.\, Rev.} \,  {\bf 123} (1961) 1053; \\
~A. ~Martin, {\it``Scattering Theory: Unitarity, Analitysity and Crossing."}
Lecture Notes in Physics, Springer-Verlag,  Berlin-Heidelberg-New-York,
1969.




\bibitem{AdS-CFT}
 J.~M.~Maldacena,
  Adv.\ Theor.\ Math.\ Phys.\  {\bf 2} (1998) 231
  [Int.\ J.\ Theor.\ Phys.\  {\bf 38} (1999) 1113]
  [arXiv:hep-th/9711200];\,\,\,
S.~S.~Gubser, I.~R.~Klebanov and A.~M.~Polyakov,
  Phys.\ Lett.\  B {\bf 428} (1998) 105
  [arXiv:hep-th/9802109];\,\,\,
E.~Witten,
  Adv.\ Theor.\ Math.\ Phys.\  {\bf 2} (1998) 505
  [arXiv:hep-th/9803131].

\bibitem{GLR}
L. V. Gribov, E. M. Levin and M. G. Ryskin,
Phys. Rep. {\bf 100} (1983) 1.


\bibitem{MUQI}
A. H. Mueller and J. Qiu,
Nucl. Phys. {\bf B268} (1986) 427.

\bibitem{MV}
L. McLerran and R. Venugopalan,
Phys. Rev. {\bf D49} (1994) 2233, 3352; {\bf D50} (1994) 2225;
{\bf D53} (1996) 458;\\ {\bf D59} (1999) 09400.

\bibitem{MUCD}
 A.~H.~Mueller,
  Nucl.\ Phys.\  B {\bf 415}, 373 (1994);
  Nucl.\ Phys.\  B {\bf 437} (1995) 107
  [arXiv:hep-ph/9408245].

\bibitem{BK}
I.~Balitsky,
[arXiv:hep-ph/9509348];\,\,
{\it Phys.\ Rev.} {\bf D60}, 014020 (1999)
[arXiv:hep-ph/9812311];\,\,\,\,
Y.~V.~Kovchegov,
{\it Phys.\ Rev.}  {\bf D60}, 034008  (1999),
[arXiv:hep-ph/9901281].


\bibitem{JIMWLK}
~J.~Jalilian-Marian, A.~Kovner, A.~Leonidov and H.~Weigert,
{\it  Phys.\ Rev.}\,  {\bf D59}, 014014 (1999),
[arXiv:hep-ph/9706377];\,\,  {\it Nucl.\ Phys.}\,{\bf B504}, 415
(1997),
[arXiv:hep-ph/9701284]; \,\,\,
J.~Jalilian-Marian, A.~Kovner and H.~Weigert,
  {\it Phys.\ Rev.}  {\bf D59}, 014015 (1999),
  [arXiv:hep-ph/9709432];\,\,\,
 A.~Kovner, J.~G.~Milhano and H.~Weigert,
 {\it  Phys.\ Rev.}  {\bf D62}, 114005 (2000),
  [arXiv:hep-ph/0004014]\,; \,\,\,
E.~Iancu, A.~Leonidov and L.~D.~McLerran,
{\it  Phys.\ Lett.}\,  {\bf B510}, 133 (2001);
[arXiv:hep-ph/0102009];\,\, {\it  Nucl.\ Phys.}\,  {\bf A692}, 583
(2001),
[arXiv:hep-ph/0011241];\,\,\,
E.~Ferreiro, E.~Iancu, A.~Leonidov and L.~McLerran,
 {\it  Nucl.\ Phys.}\  {\bf A703}, 489 (2002),
  [arXiv:hep-ph/0109115];\,\,\,
H.~Weigert,
{\it  Nucl.\ Phys.}  {\bf A703}, 823 (2002),
[arXiv:hep-ph/0004044].


\bibitem{REV}
Yuri V Kovchegov and Eugene Levin, {\it `` Quantum Choromodynamics at High Energies"}, Cambridge Monographs on Particle Physics, Nuclear Physics and Cosmology, Cambridge University Press, 2012 and references therein.



\bibitem{LERE}
  E.~Levin and A.~H.~Rezaeian,
  Phys.\ Rev.\ D {\bf 83} (2011) 114001
  [arXiv:1102.2385 [hep-ph]],\,\,
  AIP Conf.\ Proc.\  {\bf 1350} (2011) 243
  [arXiv:1011.3591 [hep-ph]],\,\,
  Phys.\ Rev.\ D {\bf 82} (2010) 054003
  [arXiv:1007.2430 [hep-ph]],\,\,
  Phys.\ Rev.\ D {\bf 82} (2010) 014022
  [arXiv:1005.0631 [hep-ph]];\,\,\,P.~Tribedy and R.~Venugopalan,
  Nucl.\ Phys.\ A {\bf 850} (2011) 136
   [Erratum-ibid.\ A {\bf 859} (2011) 185]
  [arXiv:1011.1895 [hep-ph]].

\bibitem{COR}
  E.~Levin and A.~H.~Rezaeian,
  Phys.\ Rev.\ D {\bf 84} (2011) 034031
  [arXiv:1105.3275 [hep-ph]];\,\,
%
 K.~Dusling and R.~Venugopalan,
  Phys.\ Rev.\ Lett.\  {\bf 108}, 262001 (2012)
  [arXiv:1201.2658 [hep-ph]], {\it ``Evidence for BFKL and saturation dynamics from di-hadron spectra at the LHC,''}
  arXiv:1210.3890 [hep-ph].
\,\,\,A.~Kovner and  M.~Lublinsky ,
  {\it ``Angular and long range rapidity correlations in particle production at high energy,''}
  arXiv:1211.1928 [hep-ph] and references therein.
\bibitem{LELU}
E. Levin and M. Lublinsky,
  Nucl.\ Phys.\  A {\bf 763} (2005) 172
  [arXiv:hep-ph/0501173];\,\,
  Phys.\ Lett.\  B {\bf 607} (2005) 131
  [arXiv:hep-ph/0411121];\,\, 
  Nucl.\ Phys.\  A {\bf 730} (2004) 191
  [arXiv:hep-ph/0308279].

\bibitem{BRN}
M. A. Braun,
Phys. Lett. {\bf B632} (2006) 297 [arXiv:hep-ph/0512057]; 
Eur. Phys. J. {\bf C16} (2000) 337 [arXiv:hep-ph/0001268]; 
Phys. Lett. {\bf B483} (2000) 115 [arXiv:hep-ph/0003004]; 
Eur. Phys. J. {\bf C33} (2004) 113 [arXiv:hep-ph/0309293]; 
{\bf C6}, 321 (1999) [arXiv:hep-ph/9706373]. 
M. A. Braun and G. P. Vacca,
Eur. Phys. J. {\bf C6} (1999) 147 [arXiv:hep-ph/9711486].
\bibitem{LEPP}
E.~Levin,
  JHEP {\bf 1311} (2013) 039
  [arXiv:1308.5052 [hep-ph]].
\bibitem{MUT}
A.~H.~Mueller and D.~N.~Triantafyllopoulos,
  Nucl.\ Phys.\ B {\bf 640} (2002) 331
  [hep-ph/0205167];\,\,D.~N.~Triantafyllopoulos,
  Nucl.\ Phys.\ B {\bf 648} (2003) 293
  [hep-ph/0209121].
\bibitem{IIM}
E.~Iancu, K.~Itakura and L.~McLerran,
  Nucl.\ Phys.\ A {\bf 708} (2002) 327
  [hep-ph/0203137];

\bibitem{LT}
  E.~Levin and K.~Tuchin,
  Nucl.\ Phys.\  B {\bf 573} (2000) 833
  [arXiv:hep-ph/9908317].
\bibitem{LMP}
E. Levin, J. Miller and A. Prygarin,
  Nucl.\ Phys.\  {\bf A806 } (2008)  245,
  [arXiv:0706.2944 [hep-ph]].
\bibitem{MPSI}
A. H. Mueller and B. Patel,
Nucl. Phys. {\bf B425} (1994) 471. 
A. H. Mueller and G. P. Salam,
Nucl. Phys. {\bf B475}, (1996) 293. [arXiv:hep-ph/9605302]. 
G. P. Salam,
Nucl. Phys. {\bf B461} (1996) 512; 
E. Iancu and A. H. Mueller,
Nucl. Phys. {\bf A730} (2004) 460 [arXiv:hep-ph/0308315];            
494 [arXiv:hep-ph/0309276].

\bibitem{MUSH}
 A.~H.~Mueller and A.~I.~Shoshi,
  Nucl.\ Phys.\  B {\bf 692} (2004) 175
  [arXiv:hep-ph/0402193].
\bibitem{AKLL}
  T.~Altinoluk, C.~Contreras, A.~Kovner, E.~Levin, M.~Lublinsky and A.~Shulkim,
  Int.\ J.\ Mod.\ Phys.\ Conf.\ Ser.\  {\bf 25} (2014) 1460025;
 T.~Altinoluk, A.~Kovner, E.~Levin and M.~Lublinsky,
  JHEP {\bf 1404} (2014) 075
  [arXiv:1401.7431 [hep-ph]];\,\,\,  
 T.~Altinoluk, A.~Kovner, E.~Levin and M.~Lublinsky,
  {\it ``Reggeon Field Theory for Large Pomeron Loops,''}
  arXiv:1401.7431 [hep-ph].
\bibitem{RY}
I. Gradstein and I. Ryzhik, {\it  Table of Integrals, Series, and Products},
Fifth Edition, Academic Press, London, 1994.
\bibitem{GW}
K.~J.~Golec-Biernat and M.~Wusthoff,
  Phys.\ Rev.\ D {\bf 59} (1998) 014017
  [hep-ph/9807513].


\bibitem{LR}
 E.~Levin and A.~H.~Rezaeian,
  Phys.\ Rev.\ D {\bf 82}, 014022 (2010)
  [arXiv:1005.0631 [hep-ph]].
  \bibitem{KLN}
  D.~Kharzeev and E.~Levin,
  Phys.\ Lett.\ B {\bf 523}, 79 (2001)
  [nucl-th/0108006].
\bibitem{AAQ}
  J.~L.~Albacete, N.~Armesto, J.~G.~Milhano, P.~Quiroga-Arias and C.~A.~Salgado,
  Eur.\ Phys.\ J.\ C {\bf 71} (2011) 1705
  [arXiv:1012.4408 [hep-ph]].

\bibitem{SN}
S.~Nussinov,
  {\it ``Is the Froissart bound relevant for the total pp cross section at $s=(14\,TeV)^2?$,''}
  arXiv:0805.1540 [hep-ph].



\bibitem{KOWI}
 A.~Kovner and U.~A.~Wiedemann,
  Phys.\ Rev.\ D {\bf 66}, 051502, 034031 (2002)
  [hep-ph/0112140,hep-ph/0204277];\,\,
  Phys.\ Lett.\ B {\bf 551}, 311 (2003)
  [hep-ph/0207335].
\bibitem{LLB}
 E.~Levin, L.~Lipatov and M.~Siddikov,
  Phys.\ Rev.\ D {\bf 89} (2014) 074002;\,\,\,
  [arXiv:1401.4671 [hep-ph]]; E.~Levin and S.~Tapia,
  JHEP {\bf 1307} (2013) 183
  [arXiv:1304.8022 [hep-ph]].


\bibitem{GLMMA}
  E.~Gotsman, E.~Levin and U.~Maor,
  Eur.\ Phys.\ J.\ C {\bf 71}, 1553 (2011)
  [arXiv:1010.5323 [hep-ph]].


\bibitem{KOLE}
Y.~V.~Kovchegov and E.~Levin,
  Nucl.\ Phys.\ B {\bf 577} (2000) 221
  [hep-ph/9911523].

\bibitem{PDG}
J.~Beringer et al. (Particle Data Group),
Phys. \ Rev.\ {\bf D86}, 010001 (2012)
\bibitem{DINODIF}
K.~A.~Goulianos [CDF Collaboration],
  Acta Phys.\ Polon.\ B {\bf 33} (2002) 3467
  [hep-ph/0205217] and references therein.

\bibitem{KMRREC}
  V.~A.~Khoze, A.~D.~Martin and M.~G.~Ryskin,
  arXiv:1402.2778 [hep-ph].

\bibitem{KAPO}
  A.~B.~Kaidalov and M.~G.~Poghosyan,
  {\it ``Description of soft diffraction in the framework of reggeon calculus: Predictions for LHC,''}
  arXiv:0909.5156 [hep-ph].
\bibitem{DINO}
R.~Ciesielski and K.~Goulianos,
  PoS ICHEP {\bf 2012} (2013) 301
  [arXiv:1205.1446 [hep-ph]].


\bibitem{TOTEMREC}
F. Oljemark, 15th Int.Conf. 
on Elastic and Diffractive Scattering (Saariselka) Finland, September 2013;\,\,
G. Antchev et al. (TOTEM Collaboration, Phys.Rev.Lett.{\bf 111}, 262001, (2013).
\bibitem{CMSREC}
K. Goulianos,
15th Int.Conf.on Elastic and Diffractive Scattering (Saariselka) Finland, September 2013.
\bibitem{ALICEREC}
  B.~Abelev {\it et al.}  [ALICE Collaboration],
  Eur.\ Phys.\ J.\ C {\bf 73} (2013) 2456
  [arXiv:1208.4968 [hep-ex]].
\end{thebibliography}
\end{document}